\documentclass[onecolumn,aaspp4]{aastex}
\input epsf.tex
\begin{document}
\title{Calibration of Equilibrium Tide Theory for Extrasolar Planet Systems} 
\author{Brad M. S. Hansen\altaffilmark{1}
}
\altaffiltext{1}{Department of Physics \& Astronomy and Institute of Geophysics \& Planetary Physics, University of California Los Angeles, Los Angeles, CA 90095, hansen@astro.ucla.edu}


\lefthead{Hansen }
\righthead{Two Tides}

\begin{abstract}
We provide an `effective theory' of tidal dissipation in extrasolar planet systems by empirically calibrating a model for the equilibrium tide. The
model is valid to high order in eccentricity and parameterised by two constants of bulk dissipation -- one for dissipation in the planet
and one for dissipation in the host star. We are able to consistently describe the distribution of extrasolar planetary systems in terms of period,
eccentricity and mass (with a lower limit of a Saturn mass) with this simple model. Our model is consistent with the survival of short-period
exoplanet systems, but not with the circularisation period of equal mass stellar binaries, suggesting that the latter systems experience a
higher level of dissipation than exoplanet host stars. Our model is also not consistent with the explanation of inflated planetary radii as
resulting from tidal dissipation. The paucity of short period planets around evolved A stars is explained as the result of enhanced tidal inspiral
resulting from the increase in stellar radius with evolution.

\end{abstract}


\section{Introduction}

That tidal dissipation operates in extrasolar planet systems seems beyond dispute. The eccentricities of the
closest planets are markedly smaller than those of planets with larger semi-major axes. However, the details
of how tides sculpt the distribution, and the degree to which their influence extends in determining planetary
radii is still a subject of active discussion. We have seen claims that the enhanced radii of some planets is
due to recent or ongoing tidal dissipation, while others claim not (Jackson et al. 2008, 2009; 
Miller, Fortney \& Jackson 2009; Ibgui, Burrows \& Spiegel 2010; Leconte et al. 2010). We
have also seen claims that some observed systems should be very short-lived (e.g. Hebb et al. 2010).
 Part of the problem is that the nature of tidal dissipation in
these planets is still poorly understood. Furthermore, for some systems, dissipation of tides in the star
is as important as that in the planet. Finally, much of the discussion has been couched in terms of the
tidal $Q$ parameter, a traditional measure that has been calibrated in the solar system (e.g. Goldreich \& Soter 1966)
but which is frequency-dependent, and therefore sometimes difficult to translate from one system to another.

Our goal in this paper is to exploit the growing diversity of the exoplanet parameter space to try and place
empirical constraints on the parameters of a specific tidal theory. We will adopt the equilibrium tide theory as
our basic framework, although we recognise the distinct possibility that the physics of tidal interactions is much
richer than this zeroth order treatment.
We will also allow for the possibility that the differences in stellar and planetary structure require two
different normalisations of the respective dissipative processes. Finally, we will cast our normalisations explicitly
in terms of tidal dissipation constants ($\sigma$), which are instrinsic measures of the star/planet internal structure
and viscosity, and not dependent (within the context of this theory) on the frequency of forcing. Thus, we will
be able to calculate a version of the tidal $Q$ that has a self-consistent frequency dependence.

In \S~\ref{Model} we present a brief summary of the tidal model we adopt, which we base on the model of Eggleton, Kiseleva
\& Hut (1998). In \S~\ref{Stellar} we review the literature on the related tidal dissipation problem of the circularisation
period for solar mass main sequence binaries, and what this can tell us about the dissipation in  solar mass stars. In
\S~\ref{exoplanets} we will consider the calibration by fitting to the planetary distribution, in both period, eccentricity
and planet mass. In \S~\ref{Discuss} we will discuss the implications of our results, both in terms of observations and relative
to other studies of this well-known subject.

\section{Tidal Model}
\label{Model}

Eggleton, Kiseleva \& Hut (1998) -- hereafter EKH -- present a rederivation of the equilibrium tide model of Hut (1981), which is well
suited to our purpose. It applies to arbitrarily large eccentricities (important given the observed distribution of planetary eccentricities
around other stars), and 
 and it isolates a physically motivated model of the dissipation in the planet or star, so
that we may calibrate it for a specific class of object, instead of trying to cast our calculations in terms of the oft-used but physically
obscure Q/Q' parameter\footnote{EKH demonstrate that the formalism is equivalent to the `constant time-lag' approximation often used in 
the derivation of tidal evolution equations.}.
We make the assumption that all of the host stars we discuss
have the same basic internal dissipation constant $\sigma_*$, and that all the substellar bodies (planets and brown dwarfs) have the same
$\sigma_p$. Once these quantities are specified, the mass and radii dependencies are explicit in the theory. Given this
assumption, we restrict our attention to planets with masses greater than Saturn, to guarantee that the bodies are
all of the same basic, fully-convective structure. Planets in the Neptune and SuperEarth categories may have sufficiently
different internal structures that a different dissipation constant is required.

We also fold into our definition a dimensionless factor related to the structure of the object (since this scales with the same
way as the dissipation constant). Thus, our definition of $\sigma_*$ and $\sigma_p$ includes the multiplicative factor
$(Q/(1-Q))^2$, where $Q$ is the structure factor defined in EKH, and not the quality factor of Goldreich \& Soter. This is
similar to the difference between the traditional definitions of $Q$ and $Q'$, which are related by a factor determined by
the Love number of the object in question, also determined by the density structure.

For aligned orbits, and dissipation in the planet, the semi-major axis a decreases at a rate determined by 
\begin{equation}
\frac{\dot{a}}{a} = - \frac{1}{T_p} \left[ \frac{1 + 31/2 e^2 + 255/8 e^4 + 185/16 e^6 + 25/64 e^8}{(1-e^2)^{15/2}}
- \frac{\Omega_p}{\omega} \frac{ 1 + 15/2 e^2 + 45/8 e^4 + 5/16 e^6}{(1-e^2)^6} \right], \label{adot}
\end{equation}
where e is the eccentricity of the orbit, $\omega$ is the angular frequency of the orbit, and $\Omega_p$ is the
spin of the planet. The characteristic orbital decay time is
\begin{equation}
T_p = \frac{1}{9} \frac{M_p}{M_* M} \frac{a^8}{R_p^{10}} \frac{1}{\sigma_p}
\end{equation}
where $M_p$ and $M_*$ are the planet and stellar masses, $M=M_p+M_*$ is the total mass\footnote{we retain this distinction because we will use the
case $M_p=M_*$ in \S~\ref{Stellar}.}, and $R_p$ is the planetary radius. The full equation for the orbital decay includes an
equivalent contribution that results from dissipation in the star.

The eccentricity evolution is given by 
\begin{equation}
\frac{\dot{e}}{e} = - \frac{9}{2 T_p} \left[ \frac{1 + 15/4 e^2 + 15/8 e^4 + 5/64 e^6 }{(1-e^2)^{13/2}}
- \frac{11}{18} \frac{\Omega_p}{\omega} \frac{ 1 + 3/2 e^2 + 1/8 e^4}{(1-e^2)^5} \right],
\end{equation}
and the rate of change of the spin is
\begin{equation}
\frac{\dot{\Omega_p}}{\Omega_p} =  \frac{\gamma}{2 T_p} \left[ \frac{1 + 15/2 e^2 + 45/8 e^4 + 5/16 e^6 }{(1-e^2)^{13/2}}
- \frac{\Omega_p}{\omega} \frac{ 1 + 3 e^2 + 3/8 e^4}{(1-e^2)^5} \right],
\end{equation}
where $\gamma$ is the ratio of orbital angular momentum to spin angular momentum. As in the case of equation~(\ref{adot}),
each of these equations has an equivalent for dissipation in the star. The related timescale is
\begin{equation}
T_* = \frac{1}{9} \frac{M_*}{M_p M} \frac{a^8}{R_*^{10}} \frac{1}{\sigma_*}
\end{equation}

An important feature of note in these equations is the strong dependence on stellar and planetary radius. As a result,
we will need to  properly account for the evolution of these quantities, as described below. In addition, we wish to
choose appropriate normalisation constants for $\sigma_p$ and $\sigma_*$, based on the global properties of the objects
involved. These dissipation constants scale as $\propto 1/({\rm mass \times length^2 \times time})$. If we set the timescale
to be the dynamical time of the object, we find that $\sigma_0 = ( G/(M R^7))^{1/2}$. Using scalings relative to Jupiter
and the Sun for planets and stars respectively, we define $\bar{\sigma}_p$ and $\bar{\sigma}_*$ using 
\begin{eqnarray}
\sigma_p & = & 5.9 \times 10^{-54} g^{-1} cm^{-2} s^{-1} \bar{\sigma}_p \\
\sigma_* & = & 6.4 \times 10^{-59} g^{-1} cm^{-2} s^{-1} \bar{\sigma}_* 
\end{eqnarray}

Finally, we note that the above formalism applies to aligned orbits, which is a special case of the general EKH formalism. We
adopt this for the purposes of clarity, given that our principal interest is in the orbital distribution of the planets. There
are indeed interesting questions related to the alignment of orbits and stellar spin rates, but are more particularly applicable to
questions of the origin of planetary eccentricities. For the purposes of this calculation, we adopt an eccentricity distribution
as an initial condition and leave questions of origin to a later date.

\subsection{Stellar and Planetary Evolution}
\label{PMS}

The strength of the tidal force between fluid bodies is a strong function of the radii, and thus we
need to treat the evolutionary history of stellar and planetary radii correctly.
In particular, we will find in
 \S~\ref{Stellar} that the radius evolution for the star is a non-negligible contributor, including the
pre-main sequence stage. Therefore, in the calculations to follow, we use the evolutionary models of
Baraffe et al. (1998) to describe the evolution of stellar radius with age, as a function of mass. The
evolution for some representative stars is shown in Figure~\ref{Baraffemod}.

We also need to describe the evolution of the planetary radius with mass. We use the models and formalism described
in Hansen \& Barman (2007). The thermal evolution of the planet is calculated using a Henyey code, with boundary conditions
based on models for irradiated atmospheres. The result is that the rate at which the radius shrinks is slowed as the planet
experiences more irradiation, as has been described by a variety of authors (Guillot et al. 1996; Burrows et al. 2000; Baraffe et al.2003). We also include
a bulk heating contribution, to account for energy input due to tidal dissipation in the planet. This is included as a bulk heating
rate per unit mass. Figure~\ref{TR} demonstrates the various effects. The solid curve shows a planet (with no core) with appropriate mass ($1.41 M_J$)
and irradiation to serve as an analogue to WASP-12. We choose this object because it has been modelled by both Miller et al. (2009) and
Igbui et al (2009), and can thus serve as a useful baseline for comparison with the models in those papers. The figure shows the radius
(with appropriate correction for the transit radius measurement). Inspection of the two other papers shows that our results lie in between
those of Miller \& Ibgui (after correcting for the fact that the Miller model includes a 10$M_{\oplus}$ core). The dotted and dashed
lines show the radius evolution of the same planet but including the effects of tidal heating. In calculating these curves we have parameterised
our model as best we could (given the different formalisms) to match the parameters used in Miller et al. (dotted) and Ibgui et al (dashed). The
parameters were chosen to yield the same $Q'$ values at the current separation of WASP-12. We see that we reproduce the character of the Miller
result, because we incorporated a fixed floor of eccentricity ($e=0.05$), which maintains the tidal dissipation in the planet and causes runaway
heating at small separations. The dashed line lies above the solid line, indicating that tidal heating is playing a role, but
we do not reproduce the increase in radius at late times in the Ibgui model, because our model circularises the orbit much earlier, even with
the adjusted parameterisation. We discuss the reasons for this further in \S~\ref{Discuss}.

The above comparison indicates that the planetary models used in the various studies are similar, if not exact, and that most of the differences
in the final comparison of radii are likely to result from the treatment of the tidal evolution, which is the motivation for this paper.
 
\section{Normalisation of Stellar Binaries}
\label{Stellar}

We would like to calibrate the stellar dissipation constant independently of the planets by examining the circularisation of
stellar mass binaries. We follow the formalism of Meibom \& Mathieu (2005) -- hereafter MM05 --to calibrate the model. We start with a gaussian initial distribution of eccentricities, with mean value $\bar{e}=0.35$ and
dispersion $\sigma_e = 0.21$. This represents an estimate of the eccentricity distribution of long-period ($>$50~days) binaries in the Pleiades, M35, Hyades/Praesepe, M67 \& NGC~188. Initial periods are drawn from a log-normal
distribution and the evolution is calculated using the formalism in \S~\ref{Model}.
Spin periods for the stars are initially assumed to be small (30 days), and the evolution of the stellar radius is
calculated as in \S~\ref{PMS}. MM05 found that the most
robust estimate of the tidal circularisation period $P'$ was to fit a function of the form 
\begin{equation}
 e(P) = 0.35 \left( 1 - e^{0.14 (P'-P)} \right)
\end{equation}
to the final distribution (where e=0 for $P<P'$). MM05 provide measures of the circularisation period for a series
of open clusters and representative field populations, reproduced in Table~\ref{MMTab}.

Using the above procedure, we simulate the tidal evolution of a population of $1 M_{\odot}$ stars with a variety of values of 
$\bar{\sigma}_*$ and fit the same function as MM05 to the result. By comparing to the observationally determined
circularisation period, we can then infer the empirical values of $\sigma_*$ which correspond to the observations
within the context of the equilibrium tide model. The inferred values are included in Table~\ref{MMTab}. 
Figure~\ref{TP} shows the evolution of the circularisation period with age, along with that expected from our model,
adopting the best fit value of $\bar{\sigma}_* = 5.3 \times 10^{-5}$.
 We see that the equilibrium
tide model provides an excellent description of the overall trend of the circularisation period with astrophysical age.
We also see that 
 the pre-main sequence stage is essential, as asserted by Zahn \& Bouchet. The dotted line shows the evolution of the
circularisation period if radius is held fixed, and calibrated to fit the data at late times. At earlier times this
calibration dramatically underestimates the amount of circularisation. Essentially, the circularisation period is set
by the pre-main sequence dissipation for populations with ages $< 2$Gyr and only begins
to increase, due to the main sequence dissipation, at larger ages. The best fit $\bar{\sigma}_*$ is also an order
of magnitude smaller than would have been inferred without the pre-main sequence evolution.

\section{Exoplanets}
\label{exoplanets}

As has been discussed by several authors, the circularisation of planetary orbits can be affected by dissipation in
both the planet and the star. In our model, we allow for a different dissipation constant in the two. We use the
above calibration for $\bar{\sigma}_*$ and now wish to examine whether we can explain the observed exoplanets with a
single estimate for $\bar{\sigma}_p$.

One difference in philosophy between analysing the stellar binaries and the exoplanet systems is that, in the latter,
there is good reason to believe that some systems may have had circularity imposed by formation in a disk. As such,
modelling the full eccentricity distribution is problematic without a convincing model for the underlying original 
functional form. Therefore, we proceed with a two pronged approach. First we will
analyse the exoplanet distribution in the same manner as the stellar binaries in
\S~\ref{Stellar}. This yields an initial ballpark estimate. To refine this, we will then
repeat the exercise by modelling the individual systems that define the upper envelope of
the period-eccentricity relation. The exoplanet distribution is now sufficiently well sampled that this is
a reasonably well-defined locus, and the accumulated data from fitting the individual systems should
indicate the appropriate value of $\bar{\sigma}_p$.

\subsection{Jupiters}
\label{Jups}

To begin our treatment of the exoplanet sample, we want to define a sample of exoplanets in the same spirit as those of \S~\ref{Stellar}.
 There are several pitfalls in such an analysis. The exoplanet sample covers a range of ages and stellar populations, and
is compiled through a variety of search techniques with different biases. Nevertheless, a global analysis will at least indicate a
initial guess for the dissipation rate, that can set later refinements in context.

We restrict our initial calculation to planets with a factor of three of Jupiter's mass, i.e. 0.3--3 $M_J$, orbiting
stars of $1 M_{\odot}$. The lower bound
serves to restrict our analysis to objects that share the same basic internal structure of a fully convective, H/He mixture.
The upper bound is somewhat arbitrary, but the observations of eccentricity versus mass suggest that this lies close to a
qualitative difference in behaviour. Figure~\ref{2Panel} shows the period-eccentricity relation for this sample, in a comparison
with the M35 solar mass sample of MM05. Clearly we expect a shorter circularisation period in the case of exoplanets.

We use the same initial distributions as in \S~\ref{Stellar}, except for the initial eccentricity distribution, and describe the
evolution of stellar and planetary radii using the formalism of \S~\ref{PMS}.
 Following
in the spirit of the stellar sample, we compile the distribution
function of eccentricities for planets in the above mass range by using those systems with orbital periods from 50 to 1000~days. This is fit by a gaussian,
with a somewhat lower mean value (0.2) than for the stellar sample, and with a slightly higher dispersion (0.25). We fit the resulting 
tidally evolved population with a function
\begin{equation}
 e(P) = 0.2 \left( 1 - e^{0.65 (P'-P)} \right),
\end{equation}
that is similar in spirit to the one of MM05, but with a sharper rise as dictated by the population that evolves from the model. Fitting
a similar function to the data yields a circularisation period of 2.85~days. We refer the reader again to the discussion in MM05 regarding the
choice of this function, which is designed to trace the mean trend of eccentricity with period, not the largest circular orbit.

We add one additional step to the comparison over the procedure followed in MM05. We restrict our fitting function to systems with
$e>0.02$. This was necessary because the planet sample has a significant eccess of circular orbits above the circularisation period over that expected from a model
population drawn from a tidally evolved model population. In principle, this could indicate a real eccess, suggesting that some planets do
migrate inwards on circular orbits, but could also be significantly affected by the selection effects associated with radial velocity and
transit surveys. Certainly there is no circular excess in the range of orbital periods $>10$~days. Thus, we restrict our model comparison
to the distribution of non-circular orbits.

Performing the model comparison, for a mean sample age of 3.0~Gyr, we obtain $\bar{\sigma}_p = 6.8 \times 10^{-7}$.
Figure~\ref{Pe3} shows the comparison between the observed distribution and the model population. This provides us with an approximate
mean estimate, but better constraints may be obtained by comparing to individual objects with well-determined parameters. To this end,
we can consider several specific objects whose properties define the approximate envelope of the P-e relation.

\subsection{Individual Objects}

To refine the constraints on $\bar{\sigma}_p$, we wish to examine individual systems. By modelling those that define the envelope of the period-eccentricity
relation, and by addressing the particular system parameters (stellar mass, age, etc), we hope to refine our initial estimate. In contrast to the calculations
of \S~\ref{Jups}, we perform the following calculations for the values of stellar and planetary mass specific to each system. We furthermore assume the stars
have a stellar rotation of 30~days period initially, and discuss the effects of and on stellar spin in \S~\ref{Spin!}.

\subsubsection{WASP-17b}

The shortest period planet with claimed eccentricity $>0.1$ is WASP-17b (Anderson et al. 2010). This planet has a mass of half that of Jupiter and an age estimated
to be $3^{+0.9}_{-2.6}$~Gyr. The eccentricity is $0.13^{+0.11}_{-0.07}$. The allowed values of $\bar{\sigma}_p$ depend somewhat on the assumed
age. If the age lies at the low end of the stated range, there is not much time to circularise all possible orbits, and dissipation
rates as high as $\bar{\sigma}_p = 1.2 \times 10^{-6}$ allow an eccentricity of this magnitude to survive. At the upper end of the age range, this dissipation
rate is unacceptably large and would circularise the orbit completely. In this case an upper limit of $\bar{\sigma}_p = 6.8 \times 10^{-8}$ is determined. Furthermore, with an abundance of planets in the period range, it is likely that eccentricities much larger than 0.24 are also
disallowed. This can be used to place a lower limit on the dissipation rate as well, leading to an overall constraint (including age uncertainty),
$$
6.8 \times 10^{-8} < \bar{\sigma}_p < 1.2 \times 10^{-6}.
$$

Thus, WASP-17b alone allows an  order of magnitude variation in the estimated dissipation, driven mostly by the uncertainty in the age.

\subsubsection{BD-10~3166}

This planet has a mass similar to WASP-17b 
 and  a slightly smaller semi-major axis. The measured eccentricity (Butler et al. 2000) 
is $e=0.07 \pm 0.05$, so we consider the possibility that the eccentricity is as large as 0.12.
The age determined from chromospheric activity (Saffe, Gomez \& Chavero 2005) ranges from 0.5--4.2~Gyr,
depending on the calibration used. Furthermore, the claimed radius of 1.7$R_{\odot}$ is rather large
for a 1 $M_{\odot}$ star, suggesting that the age could be closer to $\sim 10$Gyr.

If the system is $\sim 10$~Gyr old, then we can use the 1$\sigma$ lower limit on the eccentricity to
constrain $\bar{\sigma}_p < 1.7 \times 10^{-6}$. Dissipation stronger than this would have circularised the orbit over
this time span. Similarly, a lower limit can be obtained by requiring that the $1 \sigma$ upper bound
on the eccentricity match the simulations at the shortest estimated age (0.5~Gyr). Once again, this leaves
us with an order of magnitude uncertainty, dictated by the uncertain age of the star.
$$
3.4 \times 10^{-7} < \bar{\sigma}_p < 1.7 \times 10^{-6}.
$$
Another possible constraint in the same period range is possible from the system WASP-6. In this case the
error bars on the eccentricity are smaller, but the age uncertainty is even larger, and the dissipation constraints
are not markedly better.

\subsubsection{Kepler-6b}

Clearly, one of the big uncertainties in this estimate is the time that tides have had to exert their influence in
these systems.
Planets discovered by Kepler will be particularly interesting for this purpose, because their asteroseismological analyses
can potentially constrain the ages much better. Kepler has yet to announce a planet with a measureably eccentric orbit, but
we can get an estimate of what constraints might be possible by considering the system Kepler-6b.
 In the fits to the radial velocity data for Kepler-6b, the orbit was assumed to be circular (Dunham et al. 2010),
and the data appear consistent with this assumption. If we adopt an illustrative constraint
of $e<0.05$ and the calculated age of $3.8\pm 1$~Gyr for this 0.67$M_J$ planet, we obtain a lower limit 
$$
\bar{\sigma}_p > 3.4 \times 10^{-7}
$$
for the tidal dissipation. This is the amount required to circularise the planet to $e<0.05$ within 5~Gyr. As Kepler
accumulates more data, a better calibration is likely.

\subsubsection{COROT-5}

The system COROT-5b offers an improvement, by virtue of a reasonably constrained spectroscopic age (Rauer et al. 2009) of
$6.9 \pm 1.4 Gyr$. The planet is once again roughly half a Jupiter mass around a solar mass star. The eccentricity
is $e=0.09^{+0.09}_{-0.04}$, at an orbital period of 4.03 days. The range of $\bar{\sigma}_p$ compatible with this system's
error bars and age range is
$$
1.2 \times 10^{-7} < \bar{\sigma}_p < 6.8 \times 10^{-7}.
$$

\subsubsection{HD118203b}

Tracing the upper envelope of the eccentricity-period relation, we find the system HD~118203b, which has a mass of 2.13~$M_J$, a period of 6.1~days and
an eccentricity $0.31\pm 0.01$ (Da Silva et al. 2005). The age of the system is estimated to be $4.6\pm 0.8$~Gyr (Da Silva et al. 2010). In order for such a planet to possess this level of 
eccentricity at these ages, it places an upper limit on the level of dissipation of
$$
\bar{\sigma}_p < 3.4 \times 10^{-6}.
$$
This is obtained by requiring that the envelope of the P-e relation intersect the observed value at an age of 3.8~Gyr.
In principle, if this object traces the upper edge of the eccentricity distribution, we could also place a lower limit. However, the envelope
does not move significantly up even if $\bar{\sigma}_p$ is reduced to zero, because stellar tides become more important as the planet mass
increases. We will discuss this further below.

\subsubsection{HD185269}

This is a planet orbiting a subgiant (Johnson et al. 2006), with an eccentricity of $0.3 \pm 0.04$. Johnson et al. quote an age (4.2~Gyr)
with no error bar, so we estimate an error bar of $\pm 0.3$~Gyr on the basis of the error in the estimated mass and radius, and our stellar
models. The subgiant nature of this star leads to some interesting tidal behaviour. For the maximum estimated age (4.2~Gyr), we find that
we cannot match the observed parameters, even if $\bar{\sigma}_p=0$. This is because dissipation in the star is strong enough to circularise
orbits at this separation. However, at the lower end of the age range (3.9~Gyr), the stellar tides are sufficiently weakened that values
as large as $\bar{\sigma}_p = 5.1 \times 10^{-6}$ are consistent with the data. This dramatic change is because the radius of the star is increasing
as it evolves towards the giant branch, making the stellar tides increase in strength. The end result is that the constraint is much the same
as before
$$
\bar{\sigma}_p < 5.1 \times 10^{-6}.
$$

\subsubsection{Synthesis}
\label{Synth}

Fitting each of these systems yields a range of values consistent with the observed parameters.
It is possible to achieve a successful synthesis of these various constraints with  a value
$$
\bar{\sigma}_p = 5 \pm 2 \times 10^{-7}
$$
and 
$$
\sigma_* = 1.3 \times 10^{-5}.
$$

It is encouraging that $\bar{\sigma}_p$ derived in this manner is similar to that from the 
fit of the `average' model to the global
distribution. The consistency suggests that neither approach is overly bedeviled by systematic
issues related to, for example, the age of the system in question.

\subsection{Massive Planets}
\label{Massive}

In \S~\ref{Jups} we restricted our attention to planets less massive than $3 M_J$, and calibrated the planetary dissipation from the
observed distribution. There
 is potentially also interesting information in
the more massive planets, as shown in Figure~\ref{Pe4}. In the lower panel, we show the period-eccentricity distribution of the Jupiter mass sample in
Figure~\ref{Pe3}. In the upper panel we show the same relation for planets with $M> 3 M_J$. The upper envelope of the distribution appears shifted
to lower periods in the latter sample, suggesting a useful test of our equilibrium tide model. Can we explain this shift with our calibration?

\subsubsection{WASP-14b}
The planet WASP-14b (Joshi et al. 2008)  would appear to be a particularly useful system, as it has a measured eccentricity of $0.090 \pm 0.003$
(recently confirmed by Husnoo et al. 2010) at an orbital 
period of 2.24~days -- at which lower mass planets are definitely circularised. The planet mass is 7.73~$M_J$ and the estimated age of the
system is $0.75 \pm 0.25$~Gyr. 

When we adopt our calibrated model, it fails spectacularly to match the observed values of WASP-14b. The observed parameters of WASP-14b lie well above
and to the left of the allowed period-eccentricity region, for systems with the mass and age observed.
Reducing the amount of dissipation in the planet (even to zero) does not change the
envelope significantly, so it is not our assumption that dissipation is the same across all planet masses that is the problem. Rather, it is the dissipation in
the star that has to be reduced. The larger mass planet raises stronger tides on the star and this causes the stellar tides to drag the planet in further. 
A planet with the mass, period and eccentricity of WASP-14b should have been swallowed by the star on a timescale shorter than the estimated age of the
system.

\subsubsection{XO-3b}
To further illustrate the problem, let us consider the system XO-3b (Johns-Krull et al. 2008; Winn et al. 2008).
This planet has a mass of 11.8$M_J$, which is almost large enough to be a brown dwarf. With an orbital period of only 3.2~days, it nevertheless
has an eccentricity of $0.26\pm 0.02$ and an estimated stellar age of $2.8^{+0.6}_{-0.8}$~Gyr. With these parameters and the nominal calibration from
lower masses, XO-3b would have to have an orbital period of 6.6~days to survive for the estimated age. As in the case of WASP-14b,
changing the planetary dissipation rate does little to alleviate the mismatch, and we are again faced with the requirement to reduce the dissipation in
the star.

\subsection{Recalibration of the stellar dissipation}

Testing our calibration of \S~\ref{Jups} on the more massive planets shows that we cannot get a consistent calibration of the equilibrium
tide model that applies to both stellar binaries and planetary systems. The reason is that the level of stellar dissipation required to match
the circularisation in stellar binaries is too strong to match the planetary systems. We will return to the implications of this is \S~\ref{Discuss},
but, for now, let us consider a recalibration of the stellar tides using just the planetary systems. 

\subsubsection{HAT-P-2b}
At longer orbital periods (5.6~days), we find the planet HAT-P-2b (Bakos et al. 2007), with mass $9.1 M_J$ and eccentricity $0.517\pm 0.033$, at
a system age of $2.7\pm 0.5$Gyr. This is perhaps the
most extreme of the known systems, and can thus serve as a useful system to recalibrate the stellar dissipation. We simulate
this system using the individual system parameters, as in previous cases, with one exception. This system is unique amongst those
considered up to this point in that it has a stellar rotation that is faster than the planetary orbital rotation, so that the stellar tidal effects
could potentially push the planet outwards rather than inwards. For consistency with our previous models, we will assume a stellar
rotation period of 30~days initially in our calculations. We return to this system in \S~\ref{Spin!} and demonstrate that the spin
turns out to not be quantitatively important.

If we set dissipation in the planet to zero, the maximum acceptable stellar dissipation is $\bar{\sigma}_* \sim 1.6 \times 10^{-7}$.
If we use our nominal planetary dissipation of $\bar{\sigma}_p \sim 5 \times 10^{-7}$, we need to reduce $\bar{\sigma}_*$ by only
a further factor of 2. This indicates that the massive planet systems are mostly sensitive to $\bar{\sigma}_*$ and only
weakly sensitive to $\bar{\sigma}_p$. Thus, we will adopt a provisional calibration of 
$$
\bar{\sigma}_p = 5 \times 10^{-7}
$$
and
$$
\sigma_* = 8 \times 10^{-8}.
$$

If we apply these numbers to the
 systems XO-3b and WASP-14b we get satisfactory fits as well, shown in Figure~\ref{4panel}.
The reduction in the strength of the stellar tide means that planets can now orbit closer to
the star without being pulled inward and swallowed.

Figure~\ref{XO3} demonstrates how the relative strength of stellar and planetary tides change with the planet mass.
The two curves show two evolutionary scenarios. The initial condition was chosen such that it would yield the correct
period, eccentricity and age for the XO-3b system, using our final parameterisation. The lower curve is calculated
using all the same parameters and starting conditions, but reducing the mass of the planet to 1$M_J$. We see that the
lower mass planet is more rapidly circularised, and at larger distances. The consequence is that the periastron for
the lower mass planet actually moves outwards as the orbit circularises, reducing the strength of the stellar tide.
On the other hand, the periastron for the real XO-3b moves inwards, because the stellar tide is dragging the planet
inwards at a rate comparable to the rate at which the planetary tide is circularising the orbit.

\subsubsection{Revisiting the Jovian mass systems}
\label{Final}

Comparing the new calibration to the same jovian-mass planets as before shows that most are consistent with
the new calibration. One system that is not is COROT-5b, which requires a slight reduction in $\bar{\sigma}_p$ for
consistency with the new value of $\bar{\sigma}_*$, yielding a final calibration
$$
\bar{\sigma}_p = 3.4 \times 10^{-7} \label{finalp}
$$
and
$$
\bar{\sigma}_* = 7.8 \times 10^{-8}. \label{final*}
$$

\subsection{Closest Planets}

Our re-calibration of the stellar dissipation results in a reduction of the strength by
a factor of 160, relative to what is necessary to explain the circularisation of binary
stars. We can also attempt to constrain the strength of tidal dissipation by examining the
survivability of very close planets. Planetary tides will act to circularise an orbit, but
reduce to zero once this has occurred. Dissipation in the star, on the other hand, acts to
transfer angular momentum from the planetary orbit to the stellar spin, and will slowly drag
a planet inwards, even after it has circularised. Thus, we can constrain $\bar{\sigma}_*$ independently
by examining the expected lifetime of known planets.

In the limit of a circular orbit, we can calculate a characteristic inspiral time
\begin{equation}
T_{in} = 8.2 \times 10^9 yrs \left( \frac{a}{0.02 AU} \right)^8 \left(\frac{R}{R_*}\right)^{-10}
\left( \frac{M_p}{M_J} \right)^{-1} \left( \frac{\bar{\sigma}_*}{7.8 \times 10^{-8}} \right)^{-1}. \label{Tin}
\end{equation}
Table~\ref{CloseTab} shows the characteristic inspiral times for all the planets interior to
0.03~AU (excluding those that orbit M dwarfs, for whom the stellar tide might be markedly different). 
We see that the shortest ages are for WASP-18b and WASP-12b, but that both are still
$>10^8$ years and consistent with the ages estimated for the systems as a whole. The only formal
conflict is for OGLE-TR-56, whose stated age is $> 2$~Gyr, but whose estimated inspiral time
is 1.1~Gyr, and so is broadly consistent given the systematics of stellar age determination for such
stars. The fact that the observed systems can survive for a reasonable time using this parameterisation
is encouraging. We note also that reducing the ages in Table~\ref{CloseTab} by a factor of 160 would
result in significant age conflicts for many of the the systems listed there -- another indication that
the calibration using the stellar binaries is too large.

It is also of interest to consider the distribution of planets with semi-major axis. Figure~\ref{ap} shows
the cumulative distribution of jovian mass planets (detected from transits only in this case), compared
to the same distribution from our simulations. The original semi-major axis distribution was distributed
logarithmically, and the original eccentricity distribution was chosen to be the same as in \S~\ref{Jups}.
We show four models. The leftmost is the distribution left if we assume there is no dissipation in the
star. The dotted histogram indicates the distribution after 0.1~Gyr, using the dissipation in equation~(\ref{final*}).
The short dashed histogram shows the same population but after 4~Gyr. Finally, the rightmost histogram shows
the distribution after 0.1~Gyr in the case where the stellar dissipation is given by the original value in \S~\ref{Stellar}.
 In all four cases, the theoretical planets were
weighted by 1/a, in a crude attempt to account for their detectability in transit surveys. This is an admittedly
very crude model, subject to question in terms of both the initial semi-major axis and eccentricity distributions,
but demonstrates how one might go about constraining $\bar{\sigma}_*$ once a better understanding of the observational
selection effects is available. It also demonstrates that the data are more consistent with a simple model incorporating
a small amount of stellar dissipation rather than no dissipation at all or that found by our stellar binary calibration.

\subsection{Neptunes}

In principle, our calibration should extend down to lower masses, if they are accurately described as cosmic abundance
H/He mixtures. However, the current transitting planet sample does not contain any planet with mass $<0.2 M_{J}$ that
has a radius $> 0.5 R_J$. Thus, the currently known Neptune-mass objects likely all have significant core fractions, which
may dramatically influence the dissipation and may change $\bar{\sigma}_p$.

\subsection{Age Dependance}

With a calibration of the tidal dissipation in hand, one can also examine how the distribution of planets evolves with
time.
 Figure~\ref{Age2} shows the period distribution of a model population of 1$M_J$ planets around a 1$M_{\odot}$ star, comparing
the orbit period at an age of 0.1~Gyr with that at 1~Gyr. The relation deviates from linear at periods $<1.5$~days, showing that
it is planets with orbital period in this range that are expected to evolve over the evolutionary course charted by extant
observations. The spread at periods $>2.5$~days is related to the amount of circularisation that takes place over the same timespan.
Another way of phrasing this relation is that planets observed in a 1~day orbit around a 0.1~Gyr old star will spiral into the
star within 1~Gyr.

\section{Discussion}
\label{Discuss}

The issue of planetary tidal evolution has a long and illustrious history, with many applications to our own solar
system (e.g. Peale 1999). The importance of tides was realised immediately upon the discovery of the first extrasolar
planets (e.g. Rasio et al. 1996; Marcy et al. 1997) because of the extremely short periods of the first planets
discovered using radial velocities. The interest in tidal effects has been unwavering ever since, especially since
the advent of transit surveys, which strongly favour the detection of short period planets. As such, our results
have relevance to many other studies (and vice versa). However, before we can compare and contrast our results
with those of others, we need to provide a translation between the traditional manner of parameterising the dissipation
using `$Q$', and our intentionally different approach.

\subsection{The $Q'$ Parameter}

As a large fraction of the literature on this subject is phrased in terms of a particular
value of the tidal dissipation parameter $Q'$, it is of interest to see how our model compares.
If we compare of our expression with an equivalent expression cast in terms of $Q'$, such as Jackson et al (2008),
we obtain an expression
\begin{equation}
Q'_p = \left(\frac{G}{M_*}\right)^{1/2} \frac{a^{3/2}}{R_p^5 \bar{\sigma}_p} = \frac{G}{\omega} \frac{1}{R_p^5 \sigma_p}
\label{qp}
\end{equation}
where $\omega$ is the orbital angular frequency. Putting our best-fit numerical values into this
yields
\begin{equation}
Q'_p = 3.0 \times 10^8 \left( \frac{a}{0.1 AU} \right)^{3/2} \left( \frac{R_p}{1 R_J} \right)^{-5}
\left( \frac{M_*}{M_{\odot}} \right)^{-1/2}. \label{qs}
\end{equation}
Note that this is no longer a constant, but has a dependence on both semi-major axis and planetary
radius (although not planetary mass). The radius dependence is quite strong and this is, in part,
the reason for the strong effect of tidal dissipation. In general, $Q'$ will drop as the planet approaches
the star and dissipation will increase.

An equivalent analysis for the dissipation in the star yields an expression for the $Q$ of the star,
\begin{equation}
Q'_* = 6.3 \times 10^8 \left( \frac{a}{0.1 AU} \right)^{3/2} \left( \frac{R_*}{1 R_{\odot}} \right)^{-5}
\left( \frac{M_*}{M_{\odot}} \right)^{-1/2}.
\end{equation}

Figure~\ref{QQ} shows the resulting values of $Q'_p$ and $Q'_*$ for the close planet sample. We
see that a large fraction of the close Jupiter systems are characterised by present day values $Q'_p \sim 3 \times 10^7$, although
the closest systems (a few are labelled in the plot) show values a little smaller. The range of values
of $Q'_*$ shows somewhat greater variation, presumably because the strong dependence on $R_*$ means that
$Q'_*$ can vary by an order of magnitude with a 60\% change in radius. 
The functional form of equations~(\ref{qp}) and (\ref{qs}) can be easily understood by analogy with
the simple harmonic oscillator description of the equilibrium tide model (Greenberg 2009).
Many authors have estimated $Q'$ values for individual systems based on various observational constraints
and model requirements (e.g. Matsumura, Takeda \& Rasio 2008; Miller et al. 2009; Ibgui et al. 2010). These
are often significantly different than the corresponding values above, suggesting that efforts like this one
to enforce consistency within the basis of a single physical model will be increasingly necessary as the
observed sample grows. 

One of the traditional cornerstones of estimating tidal dissipation in giant planets is to draw a comparison
with the constraints on the dissipation in Jupiter, inferred by assuming that the resonant configuration
of the Gallilean satellites is driven by tidal dissipation in Jupiter (Goldreich \& Soter 1966). To cast our
results in an appropriate form for this comparison, we must re-arrange terms in the formalism of \S~\ref{Model},
moving Jupiter to the role of `Star', and Io to the role of planet. Furthermore, the forcing frequency ($\Omega_J$) of this
system comes from the rotation of Jupiter, not the orbit of Io, so that we need adopt the limit of $\Omega_J > \omega$.
Once again, in this limit, we infer
\begin{equation}
Q'_J \sim \frac{1}{2} \frac{G}{\Omega_J } \frac{1}{R_J^5 \sigma_p } \sim 5 \times 10^6
\end{equation}
for a rotation period of 10 hours and $\sigma_p = 2 \times 10^{-60} g^{-1} cm^{-2} s^{-1}$. This is only slightly higher
than the usually quoted range of $Q_J \sim 10^{5}$--$10^6$. Of course, it is also possible that the Laplace resonance
of the satellites is primordial (Peale \& Lee 2002), in which case there are no constraints on $Q'_J$ from the Io system.
This value is somewhat larger than the $Q' \sim 4 \times10^4$ inferred by Lainey et al. (2009) from a long-term astrometric
monitoring program, suggesting that a stronger frequency dependance may be required at higher forcing 
frequencies\footnote{It is interesting that the factor of 50 difference between our $Q'_J$ and the observation of Lainey
could be explained if $\sigma_p$ scales quadratically, as in the Goldreich-Nicholson formalism -- assuming the conversion
is from 3 days to 10 hours.}.

\subsection{Tidal Inflation}

One of the principal motivations for our study was the recent flurry of interest in whether tidal dissipation
can contribute to the inflated radii observed for some planets.
Several authors have attempted to account for the enhanced radii of certain planets by invoking the
dissipation due to tides to heat the planet internally. Jackson et al. (2008a,b) have run the evolution of various
systems backwards and claim that, for reasonable choices of $Q$, one could explain the tidally inflated
radii of some planets as the result of residual heat left over from recent circularisation. However,
they do not treat the evolution of the planetary radius self-consistently, an important omission, as the
strength of tides can be strongly influenced by the radius. To do this correctly required forward modelling,
and calculations of this type have been recently performed (Miller et al. 2009; Igbui et al. 2010).
Miller et al. find that tidal inflation may explain some systems, but not all.
Igbui, Spiegel \& Burrows (2009) find consistent solutions
for systems such as WASP-4b and WASP-12b with values of $Q'_p \sim 10^8$ and $Q'_* \sim 3 \times 10^6 (P_{orb}/{\rm 10~days})^{-1}$,
(which amounts to $Q'_* \sim 2.5 \times 10^7$ for these two systems). One concern about such studies is that it is not clear
how the choices of $Q$ should be related between different systems. One of the motivations for this current study was to see
what a consistent formalism might yield. Another concern is that the tidal evolution equations in
the above papers were truncated at second order
in the eccentricity, even when the eccentricity was large. This approach has been (rightly) criticised recently by Leconte et al. (2010),
who use a set of equations derived from the model of Hut (1981). This study is the closest extant one to ours, both in spirit and execution
(since the EKH model is also ultimately derived from the same model as that of Hut). 

With our final calibration, tidal inflation is not a likely cause of planetary inflation. Figure~\ref{twopanel} shows the
evolution for two prominent inflated planet systems, WASP-12 and WASP-4. In each case, the initial parameters were chosen 
such that the forward evolution with the parameterised model would yield the correct semi-major axis and eccentricity at
the estimated age of the system. In both cases (and all others we investigated) the damping of the eccentricity does cause
inflation of the planet, but happens on short times ($\sim 10^7$ years), so that the planetary cooling reduces the radius to
more traditional values by the time the planet reaches it's current semi-major axis. Discrepancies are even larger for more
distant inflated systems (like Tres-4), which experience little tidal evolution in this model. The only way tidal inflation
could explain these radii is if the planetary orbit was circularised only within the last $\sim 10^8$ years, requiring a late
injection of the planet orbit. In this we are in complete agreement with Leconte et al, who find a similar disagreement with
previous studies, and ascribe it to the fact that the truncation of the tidal equations to quadratic order weakens the strength
of the tide, allowing the planet to dissipate energy at later times, and increasing the amount of tidal inflation. This is ultimately
why we cannot reproduce the evolutionary history of Igbui et al. (2010) in Figure~\ref{TR}, no matter what parameterisation we
adopt. 

Given the similar philosophy, it is of interest to compare the values of $Q'$ assumed in Leconte et al. with our calibrations.
The Leconte model assumes a constant time lag, which EKH showed is an equivalent assumption to that of a bulk dissipation constant
such as we use here. As a result, their $Q'$ values have the same frequency dependence as ours, and they have normalised them
by assuming different values at a period of 1~day. We can use equation~(\ref{qs}) to normalise our model in a similar fashion.
For our calibration, $Q'_p \sim 10^7$ and $Q'_* \sim 6 \times 10^7$ at 1 day periods (with some scatter resulting from the radius
dependence in our model). Thus, our planetary dissipation is comparable to the upper
range assumed by Leconte et al, although our stellar dissipation is an order of magnitude weaker than the range they studied.

For completeness, we should note that our inability to find a reasonable tidal inflation model is confined to the simple
case of single planet tidal evolution. Some studies (e.g. Bodenheimer et al. 2001; Mardling 2007;
 Batygin, Bodenheimer \& Laughlin 2009; Ibgui et al. 2010) invoke a third body
to perturb the eccentricity and maintain some level of tidal inflation. Figure~\ref{Floor} shows the evolution of the
WASP-12b system in our tidal model with the addition of an eccentricity floor for the planet. As the stellar tide starts
to drag in the orbit, the continued dissipation in the planet does indeed inflate the radius, and could potentially explain
WASP-12b if $e>0.02$. This is consistent with the initial claims for the system (Hebb et al. 2009; Lopez-Morales et al. 2009), but is not consistent
with more recent measurements (Campo et al. 2010; Husnoo et al. 2010).

In conclusion, we find that it is difficult to explain the inflated planetary radii within the context of an equilibrium
tide model parameterised to match the overall properties of the exoplanet distribution, especially if one uses a tidal evolution
model that treats large eccentricity systems self-consistently.

\subsection{The Lack of Hot Jupiters around subgiant A stars}

Searches for planets around more massive stars ($>1.5 M_{\odot}$) report a higher frequency of
planets, but a conspicuous lack of close-in planets (Johnson et al. 2007; Lovis \& Mayor 2007; 
Sato et al. 2008; Niedzielski et al. 2009). However, the recent report of a likely transitting 
planet in a 1.2~day orbit around the 1.5$M_{\odot}$ star HD15082 (Collier Cameron et al. 2010) suggests that
the existence of such planets is possible at least during the main sequence stage.
 The fact that the radial velocity 
searches are performed around subgiants (so that the atmospheres are slowly rotating and more
amenable to radial velocity measurement) means that one can speculate whether the close-in planets
have been swallowed as the result of the evolution of the central star. Johnson et al. (2007)
discount this possibility because the stellar radii do not approach the semi-major axes of the
hot Jupiters until much later stages. 

However, we have seen that the stellar tidal coupling can cause a planet to spiral inwards. Can
this process, with our stellar dissipation calibration, explain the paucity of hot Jupiters around
A-type subgiants (e.g. Sato et al. 2008)? Figure~\ref{Swallow} suggests that tidal effects
will contribute, at least partially, to the removal of close-in planets. 
We can determine the degree to which tides can drag in planets, as a function of stellar radius,
by integrating the tidal evolution in concert with a model for the stellar evolution, and noting
at which point a planet from a given initial orbit is swallowed. Figure~\ref{Swallow} has been
calculated by performing this calculation for 
 a series of $3 M_J$ planets\footnote{Following the claims of Bowler et al. (2010) that the average
mass of planets around these stars is higher.}, using the evolutionary histories for metal-rich stars of mass 1.5$M_{\odot}$
 and 2.5$M_{\odot}$, using the Padova models (Girardi et al. 2000). We show the results as a function
of stellar radius, as a proxy for age.
 Also shown (dotted
line) is the criterion indicating the orbital period at the surface of the star. This is the
line used by Johnson et al., and it indeed reaches orbital periods of several days only after
the star has evolved significantly. The criterion derived from tidal evolution is somewhat
stricter, suggesting that orbital periods out as far as 8 days are denuded by the time the
star reaches the subgiant stage, and periods out to $\sim 40$~days are removed by the time the
star reaches a radius of $10 R_{\odot}$. Thus, any true hot Jupiters are removed rapidly,
although it remains to be seen whether the remaining gap between 8 and 200 days is real or
simply the result of low statistics. 

After submission of this paper, Johnson et al. (2010) 
announced the discovery of HD102956b, a planet in a 6.5~day orbit around an A subgiant. This
system supports the latter answer, and lies on the edge of the tidal survival boundary, as shown 
in Figure~\ref{Swallow}.


\subsection{Spin of the Host Stars}
\label{Spin!}

The above calculations have assumed that the host stars are spinning slowly, consistent
with what is observed for the majority of systems. If the star was spinning sufficiently
rapidly, the tidal transfer of angular momentum from stellar spin to planetary orbit could
potentially drive the planet outwards (e.g. Dobbs-Dixon, Lin \& Mardling 2004). It has been
noted that many of the massive planet hosts discussed in \S~\ref{Massive} are rotating more rapidly than
the average for the sample as a whole, so we have investigated the effects of stellar spin
on our calibration. In each case we investigated a range of initial stellar spins and repeated
the calculation with our original and final values of stellar dissipation. No appreciable
change in the period-eccentricity relation was found. As an illustration, consider the case of the HAT-P-2 system
that was used to finalise the calibration in \S~\ref{Final}. This is also the only one of the systems in which 
the stellar spin is actually faster than the orbital spin, which would reverse the sign of the
tide. We repeated our tidal evolution calculations for the same distributions in period and 
eccentricity, but now also sampling uniformly an initial stellar spin period distribution from
1 to 100 days. One might wonder whether the sign reversal of the stellar-spin dependant term would
yield a consistent fit with our original, larger, stellar dissipation rate from \S~\ref{Synth}. Figure~\ref{EOM} shows
that this is not the case. In this case we plot final stellar spin against eccentricity, for all
periods $<10$~days. We see that it is not possible to match both stellar spin and planetary eccentricity
for tidally evolved systems using the stronger dissipation. Ultimately, matching the parameters of the
HAT-P-2 system required weakening the stellar tide, irrespective of the stellar spin
contribution.

Even if the spin of the star is not dynamically important, it is possible that conservation
of angular momentum would spin the star up, if it drags a planet inwards through
tidal coupling. It has been suggested (Pont 2009) that several
observed host stars are rotating anomalously rapidly, and that this may indicate a non-negligible
influence of tides on the host star itself. Within the context of our model calibration, none
of the proposed systems (HD189733, COROT-2b, HAT-P-2b and XO-3b) exhibits a sufficiently strong
coupling to spin the stars up to significant levels. The more recently discovered system, WASP-18b,
offers perhaps the best chance of observing stellar spin-up, as it is a massive planet in a very
short period orbit (Hellier et al. 2009). However, even this system, within the context of stellar solid-body rotation,
fails, by an order of magnitude, to spin the star up to the observed rotation period of 5.6~days
(assuming a system lifetime of 1.5~Gyr). Thus, we do not expect the dynamics of the previous sections
to be significantly affected by stellar rotation. It is possible, however, that the observed rotation
period represents the spin-up of the surface convection zone only. WASP-18b is an F-star and thus it
is quite possible that the inspiral spun-up the surface convection zone while leaving the interior
less rapidly rotating.

The best system for observing outward migration driven by stellar spin is WASP-33b (Collier Cameron et al. 2010).
The estimated stellar rotation period is 0.81~days, and the orbital period is 1.2~days. This is sufficiently close 
that the tidal coupling is strong enough to be important even with the calibration of \S~\ref{Final}. Figure~\ref{WASP33}
shows the expected orbital evolution of the system, assuming the surface rotation represents the bulk solid
body rotation of the star. We see that the planet is eventually driven out to orbital periods $\sim$~2~days.
The effect gets rapidly weaker with distance, and planets with orbital periods $>$3~days experience little
change.

\subsection{Internal Structure and Microphysical Dissipation}

We have calibrated the planetary and stellar dissipation using a single bulk constant, within the context of the
equilibrium tide model.
It is clearly of some
interest to understand how the final value compares to more detailed models for the true microphysical dissipation. 
The simplest microphysical model is dissipation of the equilibrium tide due to the turbulence in the convection zone.
In appendix~\ref{Internal}
we use the formalism of Eggleton et al to relate the bulk $\bar{\sigma}_p$ and $\bar{\sigma}_*$ to the internal dissipative
processes, using our models for planetary and stellar structure.

For the planets, we find that the amount of dissipation required by our calibration is several orders of magnitude lower than that
obtained if one adopts a simple local turbulent viscosity  based on convective velocities and scale heights. This
is encouraging, because the largest eddy overturn times are of the order of years, and so it is not clear
how strong a coupling is likely to occur to forcing periods on the order of days. Indeed, there are several proposed
prescriptions regarding how inefficient such a coupling is. Zahn (1989) proposes that the strength of the turbulent
viscosity is reduced by a linear factor in the period, while Goldreich \& Nicholson (1977)  propose a quadratic scaling. Our
calculations in appendix~\ref{Internal} favour a value intermediate between these prescriptions. This is perhaps not surprising in
the light of recent studies by Penev, Barranco \& Sasselov (2009), who have performed simulations of forced anelastic convection.
They find rates of dissipation that scale similar to the Zahn prescription when the forcing period is not too different from
the eddy overturn times, but faster losses of efficiency as the forcing frequency becomes significantly faster. They suggest that
the ultimate root of this behaviour is that the turbulence deviates from the Kolmogorov form used to derive the above scalings.
While our results clearly cannot shed any further light at this level of detail, it is encouraging that the order of magnitude
values are within range of what one might naturally expect.

Our estimate of the stellar dissipation is not as sensitive to the various scalings, as the eddy overturn periods are of order
days in this case, and thus should couple well to the observed properties. The effective calibration is a factor $\sim 10$ lower
than what we infer from the simple model estimate, so that it is consistent with a slightly inefficient dissipation of the nominal
equilibrium tide. The fact that this equilibrium tide value is broadly consistent with our calibration based on planets but not
based on stars finds broad support in studies such as those of Ogilvie \& Lin (2007). The essential difference is that planet host
stars do not have spins synchronised to the orbital period, while stars in close, equal-mass binaries do. In the latter case, Ogilvie
\& Lin show that the equivalence of forcing period and rotation period allows for the excitation of inertial modes (an example of
the `dynamical tide'), which can 
substantially enhance the rate of tidal dissipation. Similar modes are not excited by planets because the host stars rotate much slower
than the orbital period. Excitation of modes in the radiative core (Goodman \& Dickson 1998; Witte \& Savonije 2002; Ogilvie \& Lin 2007; Barker \& Ogilvie 2010) may also
contribute an enhanced dissipation the case of planets, but the extent to which they do depends on the uncertain non-linear dissipation
of waves that propagate to the center (in the limit of low dissipation, the excitations result in global modes with specific resonant
frequencies, which do not generate a broad-band tidal response).

\section{Conclusions}

We have used the current sample of observed exoplanets to calibrate a model for tidal evolution based on that of 
Eggleton, Kiseleva \& Hut (1998). This formalism has several advantages over those used in many other studies, in
that it can be used to treat systems with large eccentricities, and offers a specific model for the frequency dependence
of the tidal dissipation. Our calibration makes use of the distribution of planets and stars in terms of period, eccentricity
and mass. Our principal conclusions are 
\begin{itemize}
 \item We can successfully calibrate the model using the distribution of planets. By considering the variation of eccentricity
amongst planets of different mass, we are able to simultaneously constrain dissipation in both planets and the main sequence host stars.
We find that the calibration requires a level of dissipation in the host star that is inconsistent with the circularisation period
of stars in equal-mass main sequence binaries. This agrees with prior studies which find that the generation and dissipation of inertial
waves in this latter class of system may lead to enhanced dissipation relative to non-synchronous systems which characterise the exoplanet
host sample.
\item With this calibration we are unable to reproduce prior claims that the bloated radii of some hot Jupiters are the result of recent
tidal circularisation and dissipation. This disagreement is not only the result of our calibration, but also due to the fact that our
model treats high eccentricity systems with greater fidelity than many previous studies.
\item Our calibration of stellar dissipation is consistent with the existence and survival of recently discovered planets with orbital
periods $\sim 1$~day or less. We show that planets in this period regime do experience significant tidal evolution on the timescale of
Gigayears, but are expected to survive for the lifetimes of their observed hosts.
\item We find that this level of tidal inspiral may also explain the lack of planets in close binaries observed around A-type subgiants.
As the stellar radius increases, the evolving star exerts a stronger tidal force and drags planets inwards, to the extent that planets
as far out as $\sim 8$~days may be swallowed by the time the stars evolve to the point at which they are now observed.
\item We have compared our model to simple estimates of turbulent dissipation of the equilibrium tide. Our results suggest a value that
lies somewhere between the two proposed scalings for the efficiency of dissipation as a function of forcing frequency. 
\end{itemize}

These results are encouraging in the sense that it is possible to describe the global distribution of exoplanets within a simple tidal evolution
model. However, we must also recognise that the equilibrium tide model used is amongst the simplest possible descriptions of a complex physics
problem. Given the success of the simple model, it may be of future interest to follow this study with a similar study based on a more complex
microphysical description. 

\acknowledgements 
The author would like to thank Phil Armitage, Kristen Menou, David Spiegel, Jeremy Leconte and an anonymous referee
 for comments on the manuscript, and the participants
in the Kavli Institute for Theoretical Physics workshop on Extrasolar Planets for stimulating the beginnings of this project.

\newpage

\begin{deluxetable}{lccc} 
\tablecolumns{4} 
\tablewidth{0pc} 
\tablecaption{Tidal circularisation periods for solar mass binaries of various ages. 
\label{MMTab}} 
\tablehead{ 
\colhead{Population} & \colhead{$\log t_9$}   & \colhead{$P'$ (days)}  & \colhead{$10^{5} \bar{\sigma}_* $}
 }
\startdata 
PMS binaries & -2.5 & $7.1^{+1.2}_{-1.2}$ & $6 \pm 2$ \\
Pleiades     & -1.0 & $7.2^{+1.8}_{-1.9}$ & $3.1 \pm 2$ \\
M35          & -0.8 & $10.2^{+1.0}_{-1.5}$ & $9.7 \pm 4.8$ \\
Hyades/Praesepe & -0.2 & $3.2^{+1.2}_{-1.2}$  & $\cdots$ \\
M67 & 0.6 & $12.1^{+1.0}_{-1.5}$ & $7.0 \pm 3.0$ \\
NGC188 & 0.8 & $14.5^{+1.4}_{-2.2}$ & $9.7 \pm 3.8$ \\
Field & 0.95 & $10.3^{+1.5}_{-3.1}$ & $0.8 \pm 0.6$\\
Halo  & 1.0 & $15.6^{+2.3}_{-3.2}$ & $3.6 \pm 2.0$\\
\enddata 
\end{deluxetable} 

\begin{deluxetable}{lccccc}
\tablecolumns{6}
\tablewidth{0pc}
\tablecaption{Tidal Inspiral Times for Close Planets 
\label{CloseTab}}
\tablehead{
\colhead{Planet} & \colhead{$M_p (M_J)$}   & \colhead{R$_*$/R$_{\odot}$}  & \colhead{a (AU)} & \colhead{T$_{in}$} (Gyr) & \colhead{Age (Gyr)}
 }
\startdata
WASP-19b & $ 1.15 \pm 0.08 $ & $0.93 \pm 0.05$ & 0.0164 & $3.0 \pm 0.2$ & 0.6 \\
COROT-7b & $0.0151 \pm 0.0025$ & $0.93 \pm 0.03$& 0.0172 & 336$\pm$56 & 1.5 \\
WASP-18b & $10.4 \pm 0.4$ & 1.23$\pm 0.05$& 0.02047 & 0.12$\pm$0.005 & $0.63^{+0.95}_{-0.53}$ \\
OGLE-TR-56 & $1.29 \pm 0.12$ & $1.32\pm 0.06$ & 0.0225 & $1.1 \pm 0.1$ & $>2$ \\
TrES-3 & $1.92 \pm 0.23$ & 0.813$^{+0.012}_{-0.027}$ & 0.0226 & 82$\pm 10$ & $\cdots $ \\
OGLE-TR-113& $1.32 \pm 0.19 $& 0.77$\pm 0.02$ & 0.0229 & 249$\pm 36$ & $>0.7$ \\
WASP-12b & $1.41 \pm 0.1$ & $1.57 \pm 0.07$ & 0.0229 & 0.24$\pm 0.02$ & $\cdots$ \\
WASP-4b & $1.12 \pm 0.08$ & $1.15 \pm 0.28$ & 0.023 & $5.5 \pm 1.3$ & $\cdots$ \\
COROT-1b & $1.03 \pm 0.12$& 1.11$\pm 0.05$ &  0.0254 & $18 \pm 2$ &$\cdots$ \\
WASP-33b & $<4.1$ & 1.44$\pm 0.03$ & 0.0256 & $>0.36$ & 0.25\\
WASP-5b & $1.64 \pm 0.08$ & 1.08$\pm 0.04$ & 0.0273 & $27 \pm 1$ & $3 \pm 1.4$ \\
COROT-2b & $3.31 \pm 0.16$ & $0.90 \pm 0.02$ & 0.0281& 108$\pm 5$ & $\cdots$ \\
\enddata
\end{deluxetable}

\figcaption[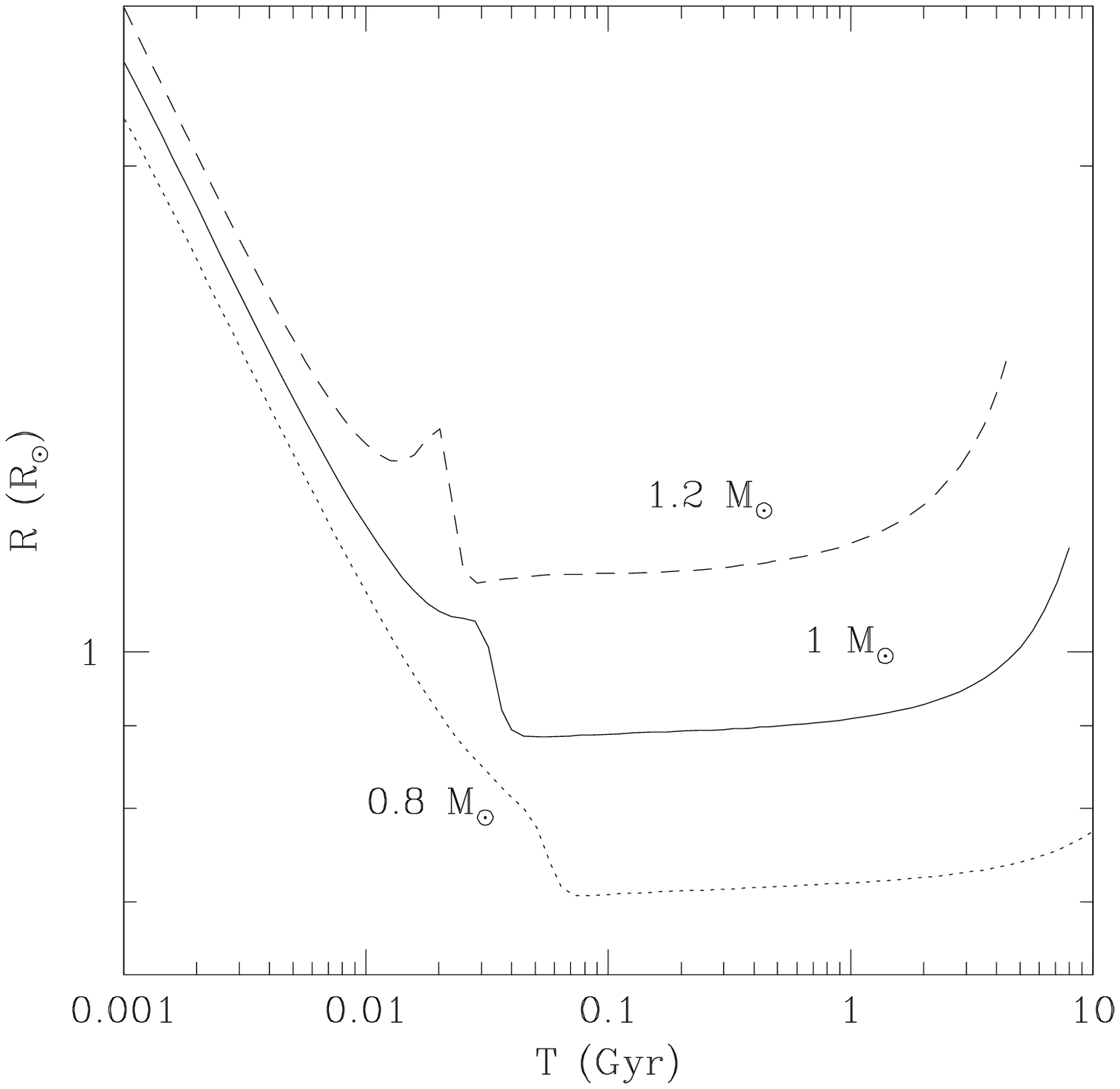]{The curves show the stellar radius evolution from the models of Baraffe et al. (1998),
for masses of 0.8, 1.0 and 1.2 $M_{\odot}$. Although the 1.2 $M_{\odot}$ star is only about 60\% larger
than the 0.8$M_{\odot}$ star over most of the age range, this can amount to a factor of 100 when raised
to the tenth power. The temporal evolution of a given mass star has an even stronger effect.
\label{Baraffemod}}

\figcaption[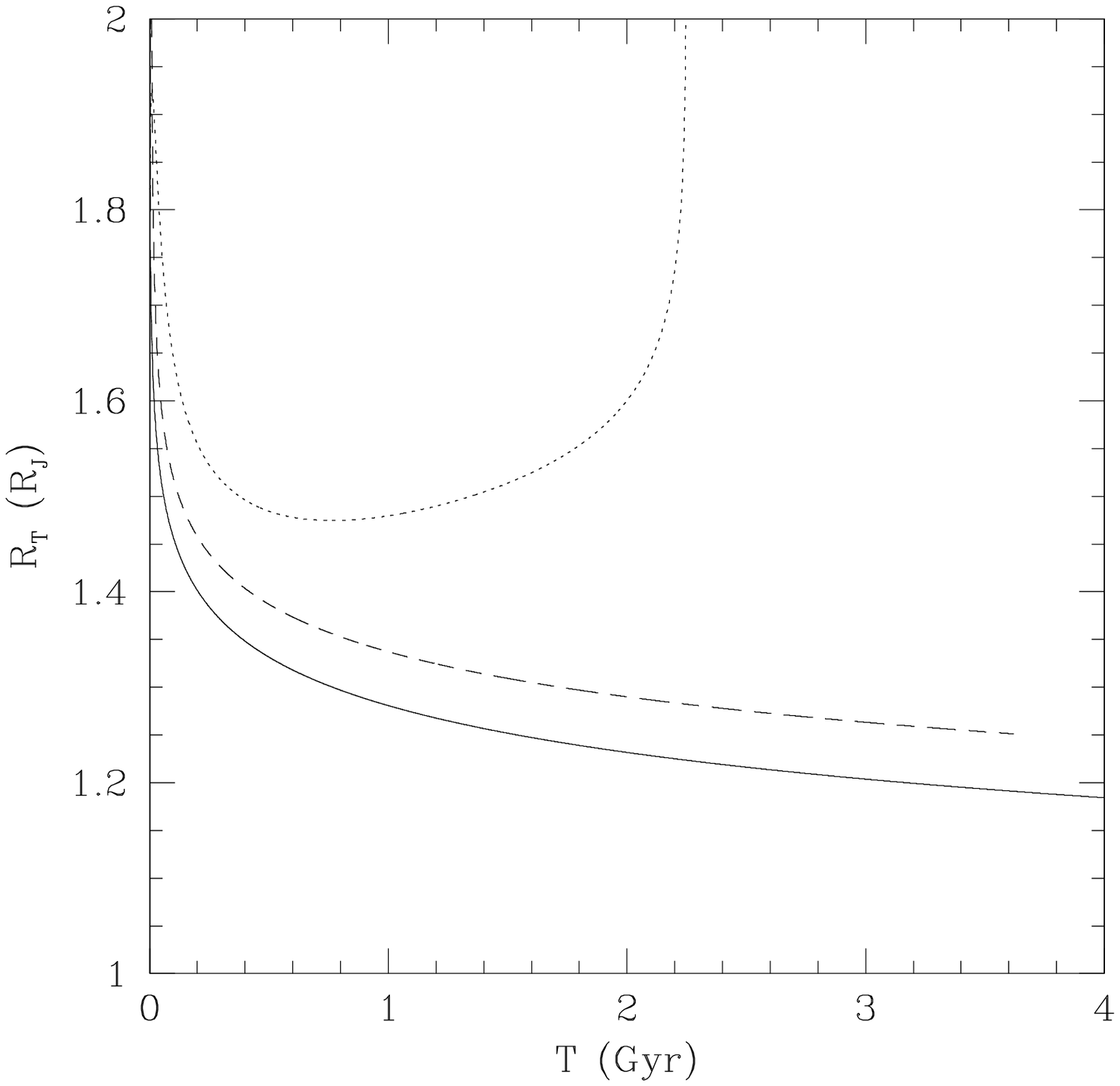]{ All three curves have been calculated for a 1.41$M_J$ planet, with an irradiation
appropriate to that of the WASP-12b system. The solid curve assumes no extra internal heating due to
tides. It can be compared with the curves in Figure 11 of Miller et al. (2009) and Figure 2 of Ibgui, Spiegel
\& Burrows (2009). The dotted and dashed curves represent attempts to match the figures of those respective
papers by parameterising our models to yield the same value of $Q_p$ and $Q_*$ at the current semi-major
axis of WASP-12. We also incorporate an eccentricity floor of 0.05 for the dotted line, which yields the
same runaway heating as used by Miller et al. Such an eccentricity was initially claimed for WASP-12, but
subsequent observations seem to belie this, so the approximation may not be valid.
\label{TR}}

\figcaption[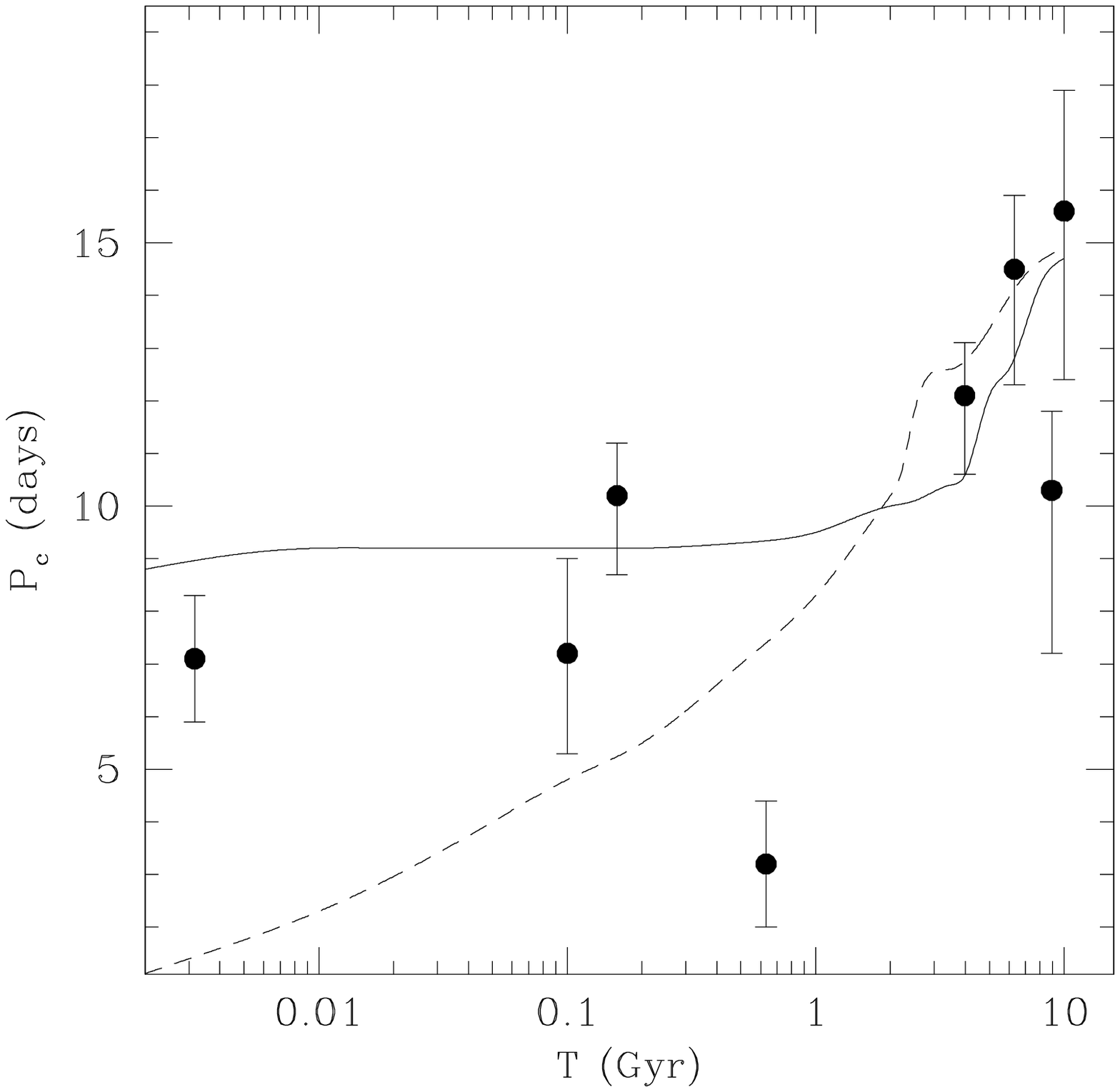]{ The points indicate the circularisation period
as a function of age for different stellar populations, from Meibom
\& Mathieu (2005). The solid line is the best-fit version of our model,
including pre-main sequence evolution. For comparison, the dotted line
is what we obtain with the best fit value of $\tau_p$ for the field
population of the local solar neighbourhood, but assuming a fixed radius of $1 R_{\odot}$. 
We see that this strongly underpredicts the circularisation of early-type binaries.
 \label{TP}}

\figcaption[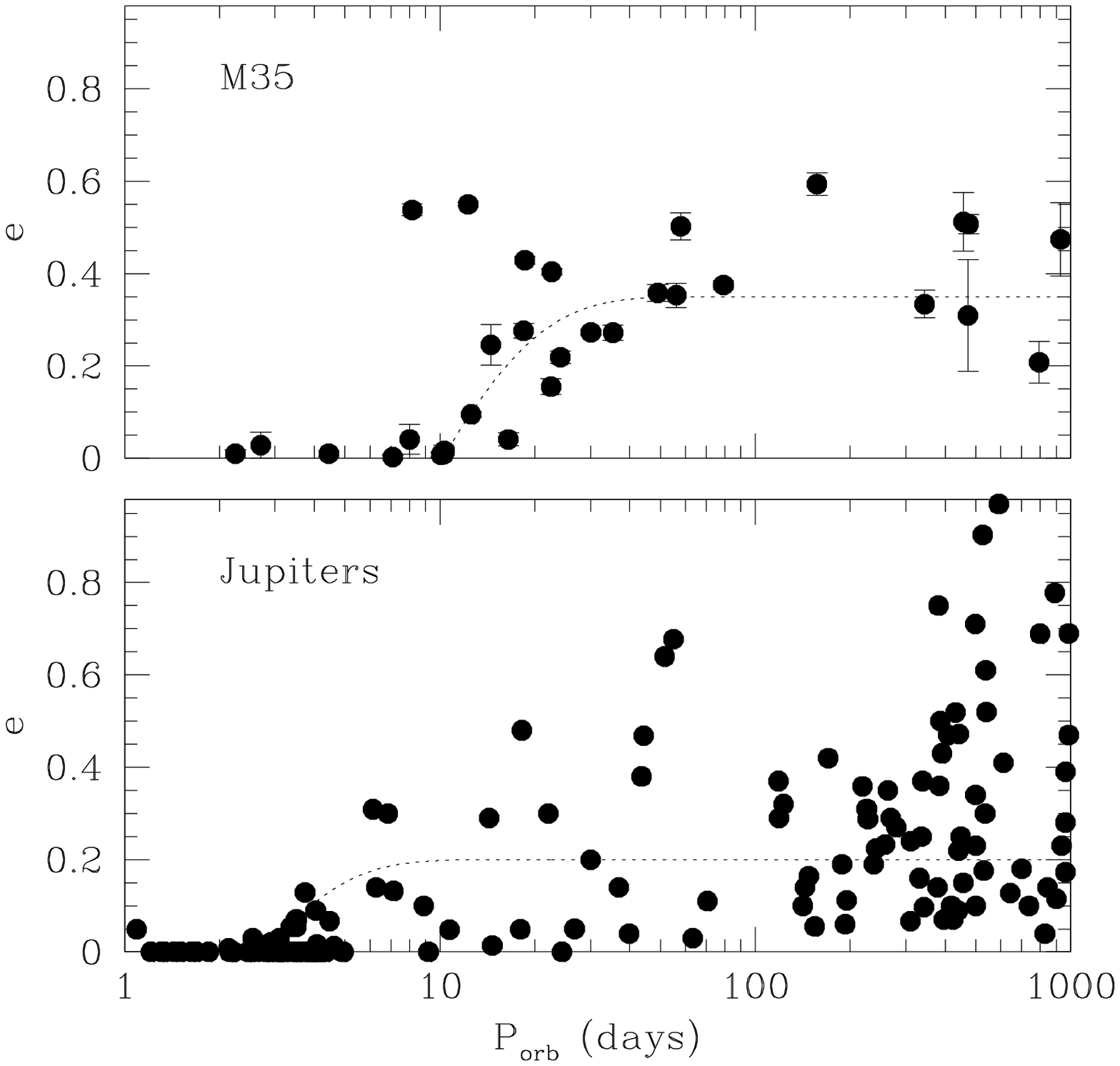]{ The top panel shows the eccentricity-period relation
for the 0.15~Gyr old open cluster M35 (Meibom \& Mathieu 2005). The lower
panel shows the eccentricity-period relation for the sample of exoplanets
with masses between $0.3 M_J$ and $3 M_J$, that orbit stars with masses
between $0.7 M_{\odot}$ and $1.5 M_{\odot}$. In each panel, we show the
best-fit circularisation function for that sample.
 \label{2Panel}}

\figcaption[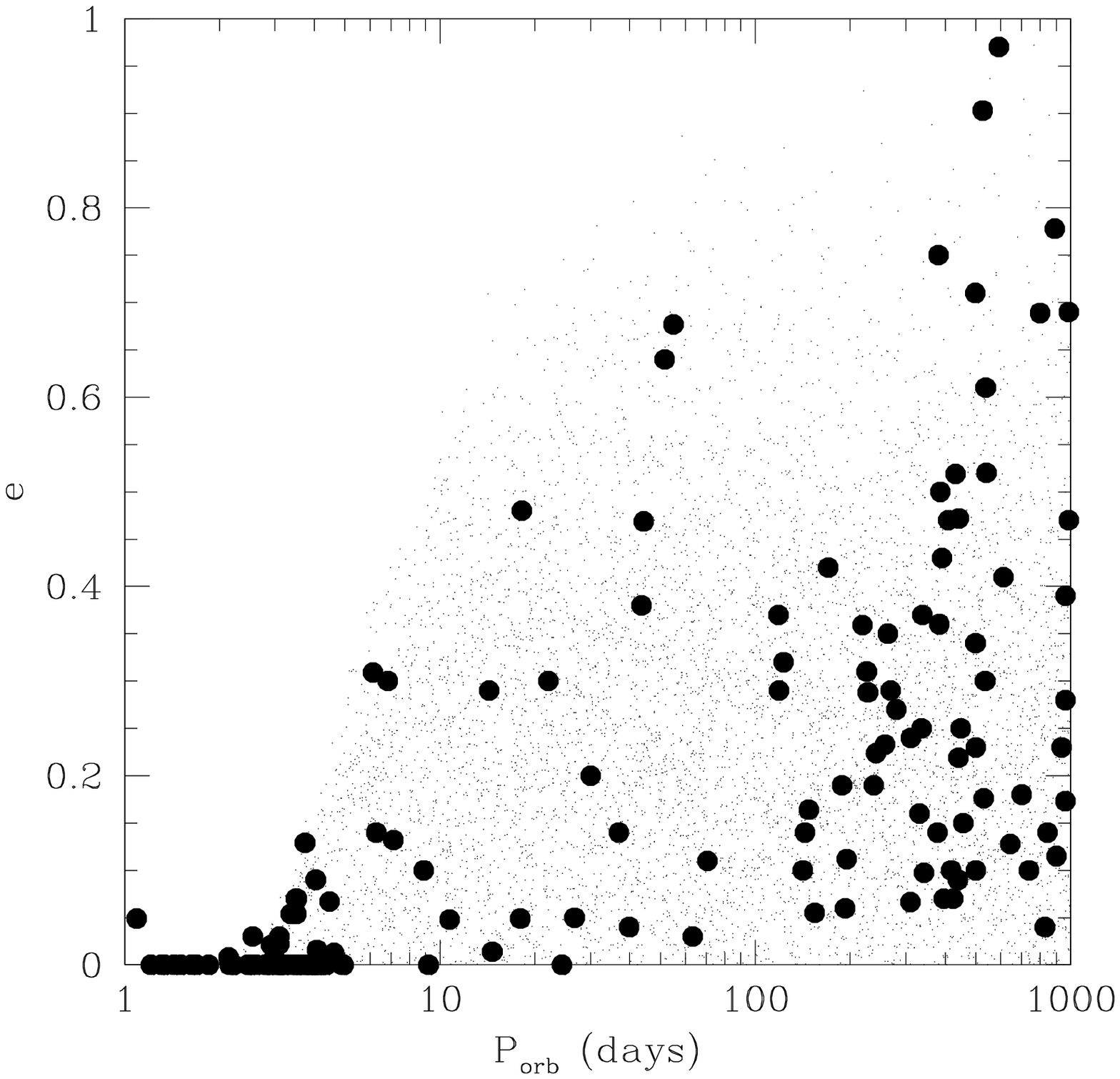]{ The solid points show the eccentricity-period relation
for Jovian-mass exoplanets. The small dots indicate a model population,
drawn from the long-period eccentricity distribution and then tidally 
evolved for 3.0~Gyr, with our best fit values for $\bar{\sigma}_p$ and $\bar{\sigma}_*$.
The model traces the envelope of the observations very well. The thinning out
of model points above $e \sim 0.5$ is the result of the original distribution,
which had a mean of 0.2 and a dispersion of 0.25, as dictated by the long period
objects.
 \label{Pe3}}

\figcaption[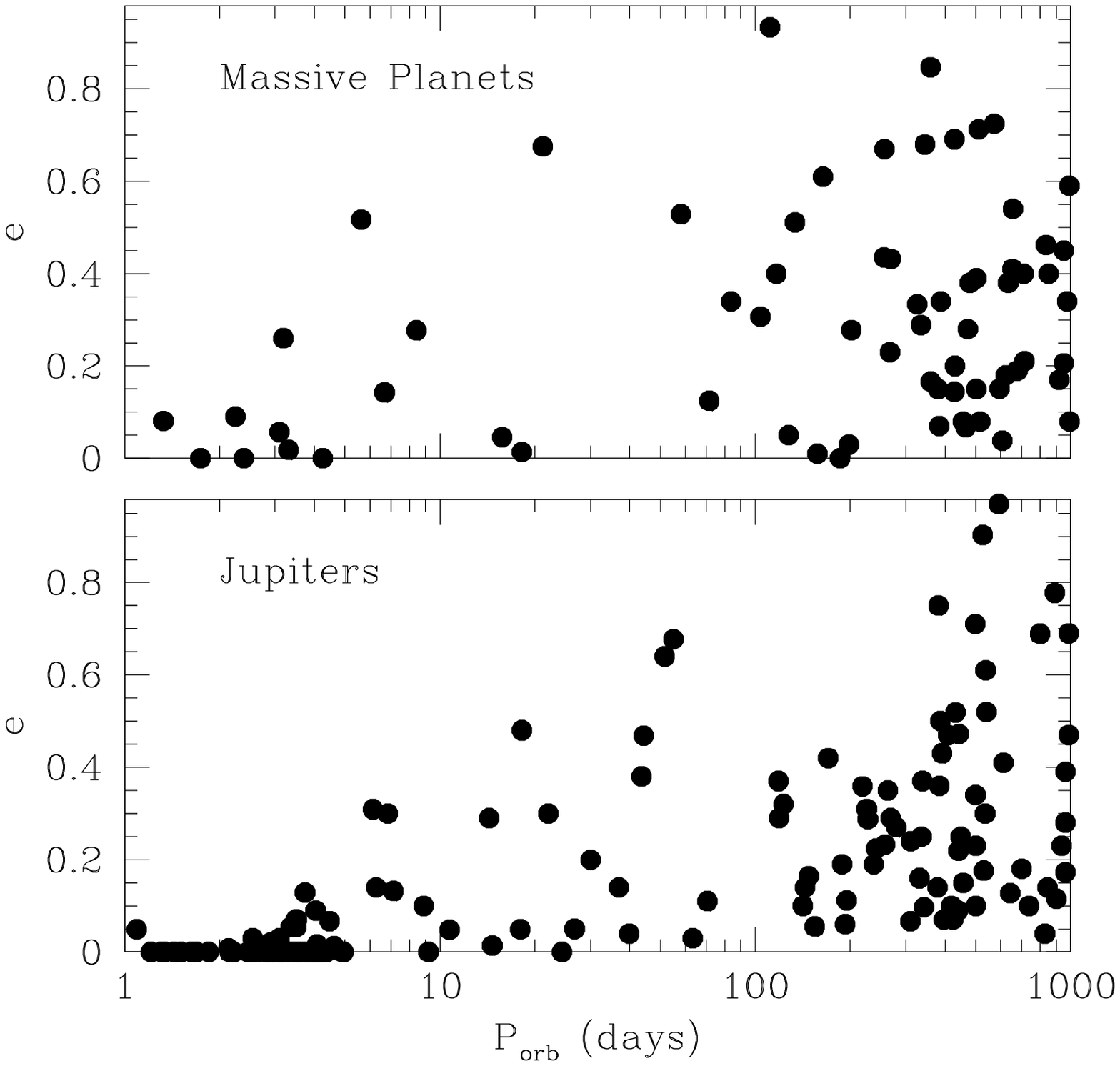]{The lower panel masses between $0.3 M_J$ and $3 M_J$, that orbit stars with masses
between $0.7 M_{\odot}$ and $1.5 M_{\odot}$. The upper panel shows the equivalent sample of
planets with masses above $3 M_J$. The more massive planets appear more eccentric.
\label{Pe4}}

\figcaption[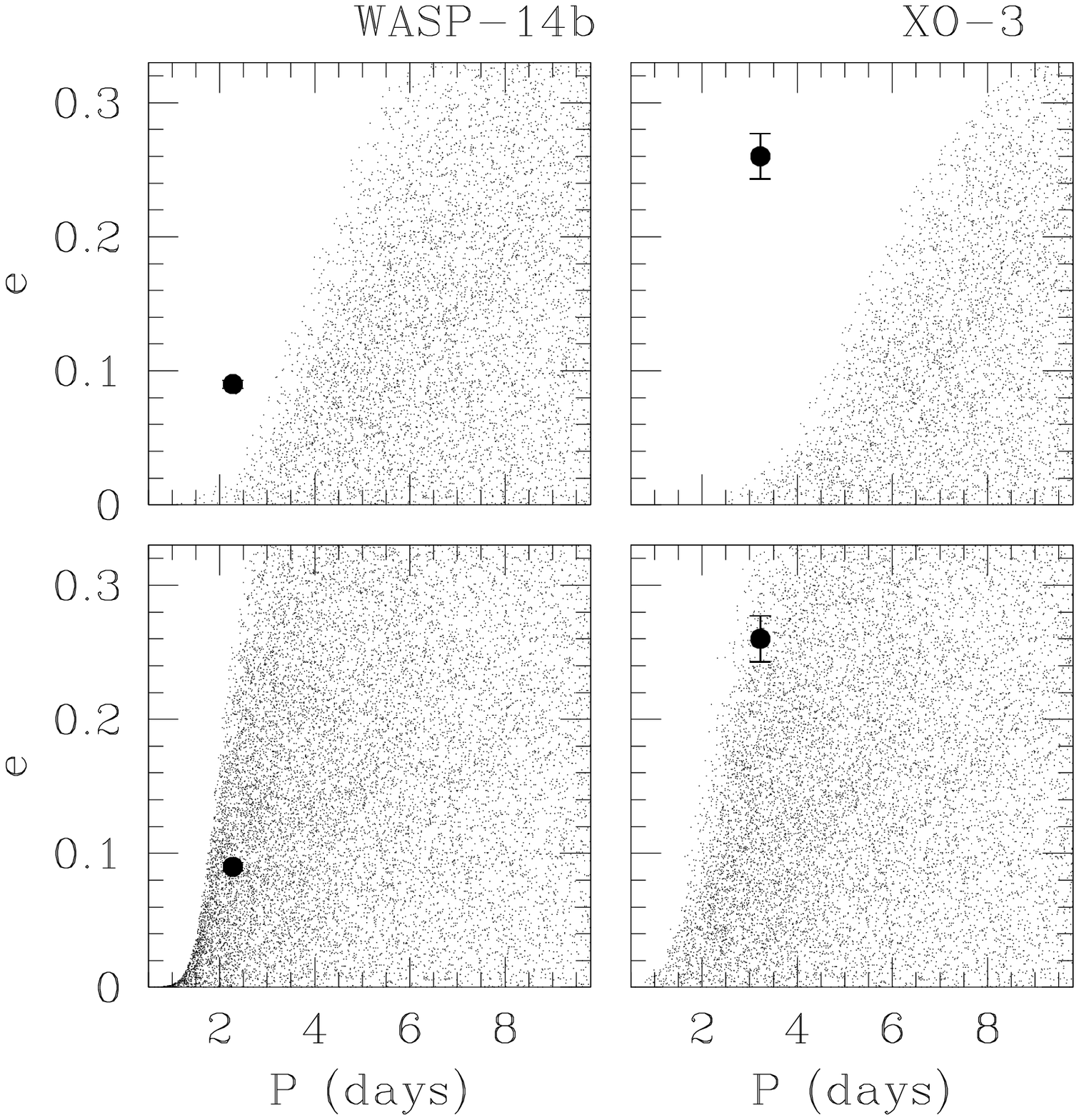]{The top two panels show the comparison between observations and model realisations
for the original parameterisation of tidal dissipation in \S~\ref{Synth}, for the cases of the massive
planet systems WASP-14b and XO-3. In the bottom two panels, we show the same comparisons, but now with
the modified calibration of \S~\ref{Final}, tuned to  fit the most extreme massive system HAT-P-2b. 
\label{4panel}}

\figcaption[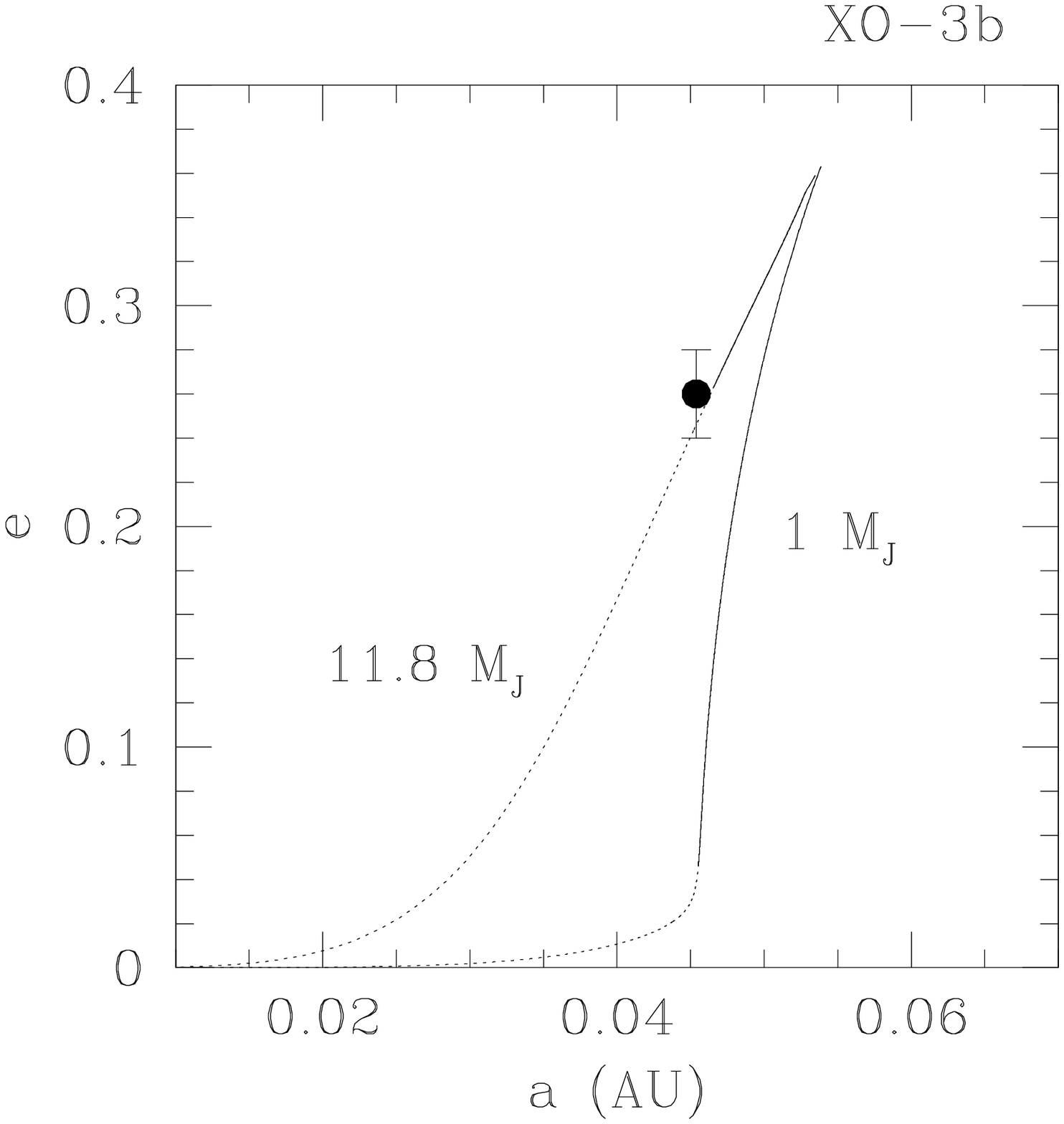]{The upper curve shows the evolution of a star-planet with the parameters of the XO-3b
system. The solid line shows the first 3~Gyr of tidal evolution using our nominal parameterisation of
the equilibrium tide model. The dotted line shows the subsequent evolution of the system.
 The lower curve shows the same evolution keeping every parameter the same, except
for the mass of the planet, which is reduced to 1~$M_J$. The divergence of these curves shows why the envelope
of the period-eccentricity relation is different between planets of different masses.
\label{XO3}}

\figcaption[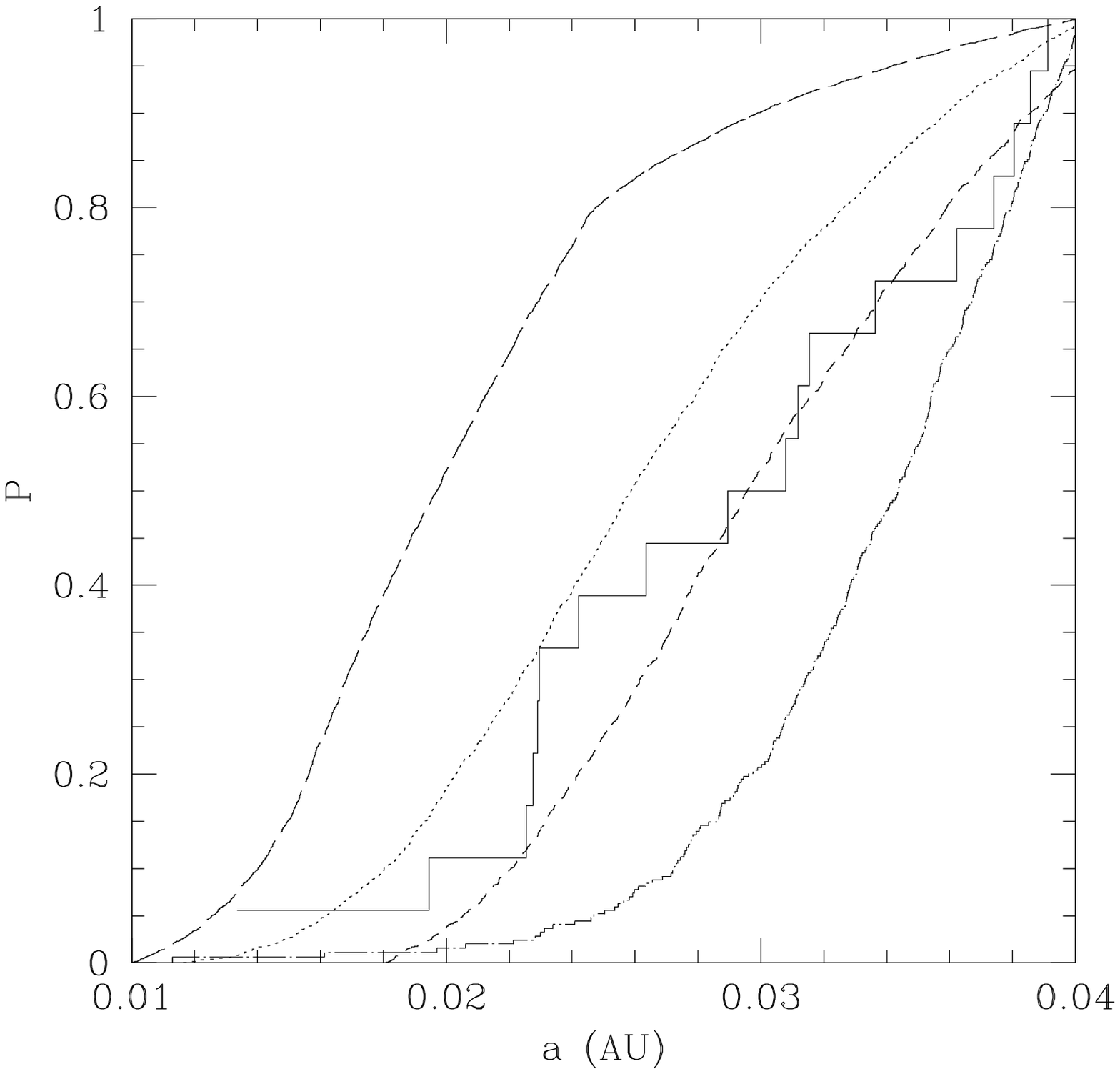]{The solid histogram shows the cumulative distribution of planets within 0.04~Au (excluding
M star systems) discovered in transit surveys. The short dashed line indicates the distribution expected from
our model with the final calibrations, at an age of 3~Gyr. The dotted line is for the same model, but with an
age of 0.1~Gyr. We see that these trace roughly the correct distribution. The upper curve is the distribution
in the model (at 3~Gyr) if we set $\bar{\sigma}_*=0$. Clearly this dramatically overpredicts the number of close
planetary systems. The lower dashed curve is recovered if we use the value of $\bar{\sigma}_*$ derived from the
stellar binaries. This clearly underpredicts the number of close planets. All model curves are weighted by
$1/a$, in order to approximately account for the detectability in a transit survey.
\label{ap}}

\figcaption[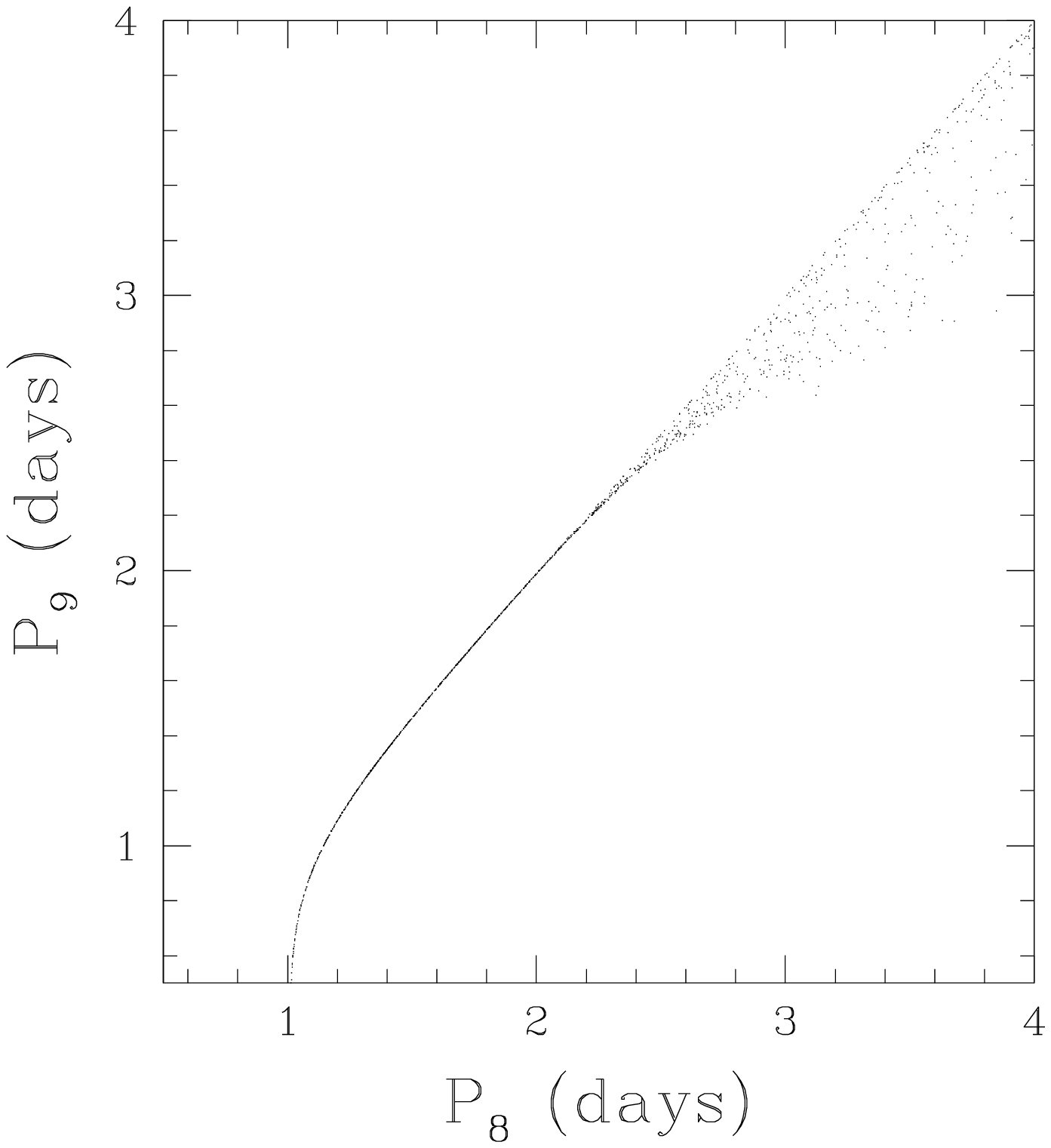]{The relation between the orbital period of a planet at an age of $10^8$~yrs ($P_8$) and
an age of $10^9$~yrs ($P_9$) indicates that systems with orbital period $<1.5$~days are expected to undergo
significant orbital evolution on astrophysical timescales. At the upper end of the plot, we see that the
degree of circularisation changes on similar timescales for periods $>$2.5~days.
\label{Age2}}

\figcaption[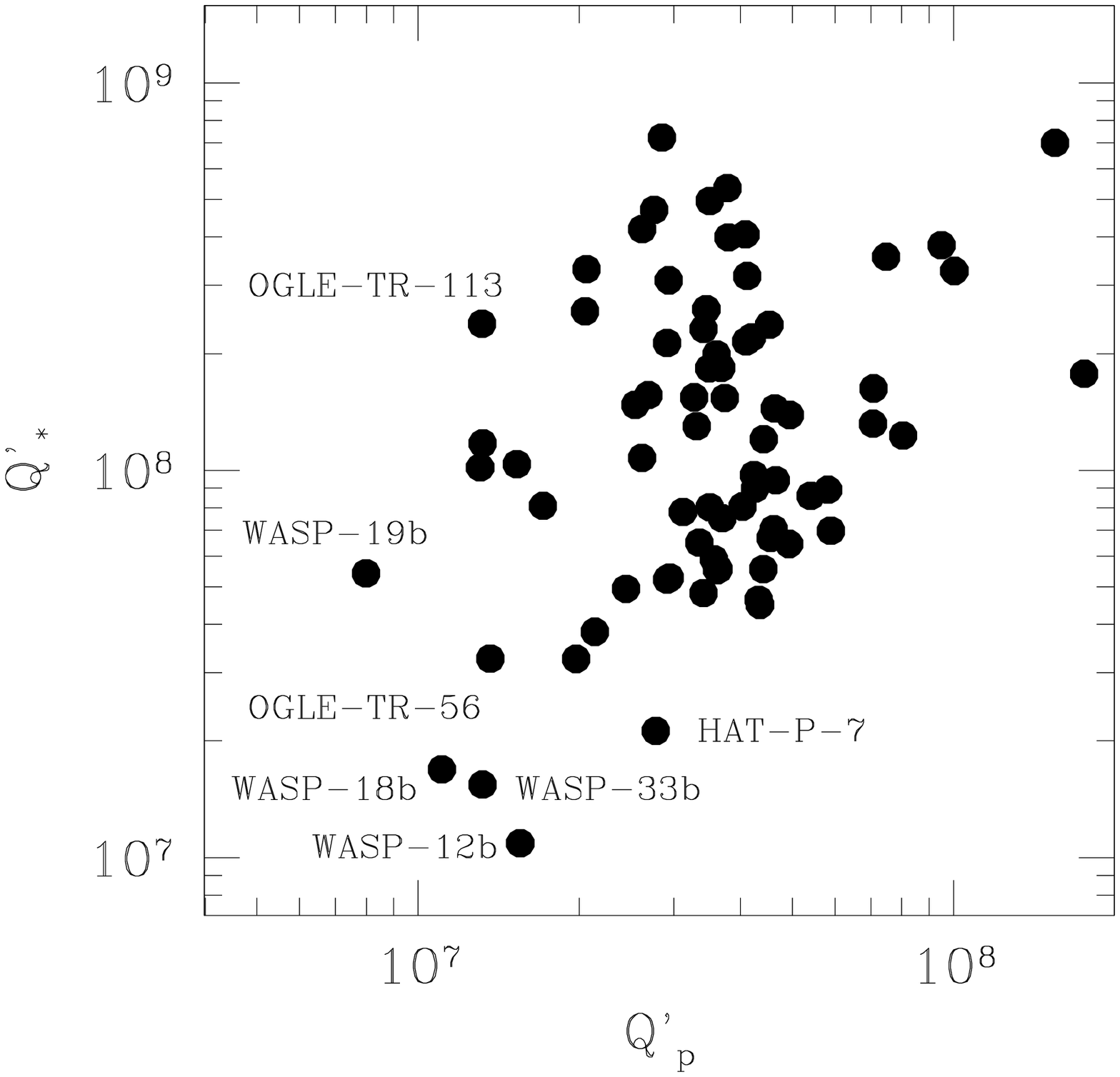]{The points show the estimated present day values of $Q'_p$ and $Q'_*$ for the
known hot-Jupiter systems, given our calibration of the equilibrium tide model. A few of the
outliers are labelled. 
\label{QQ}}

\figcaption[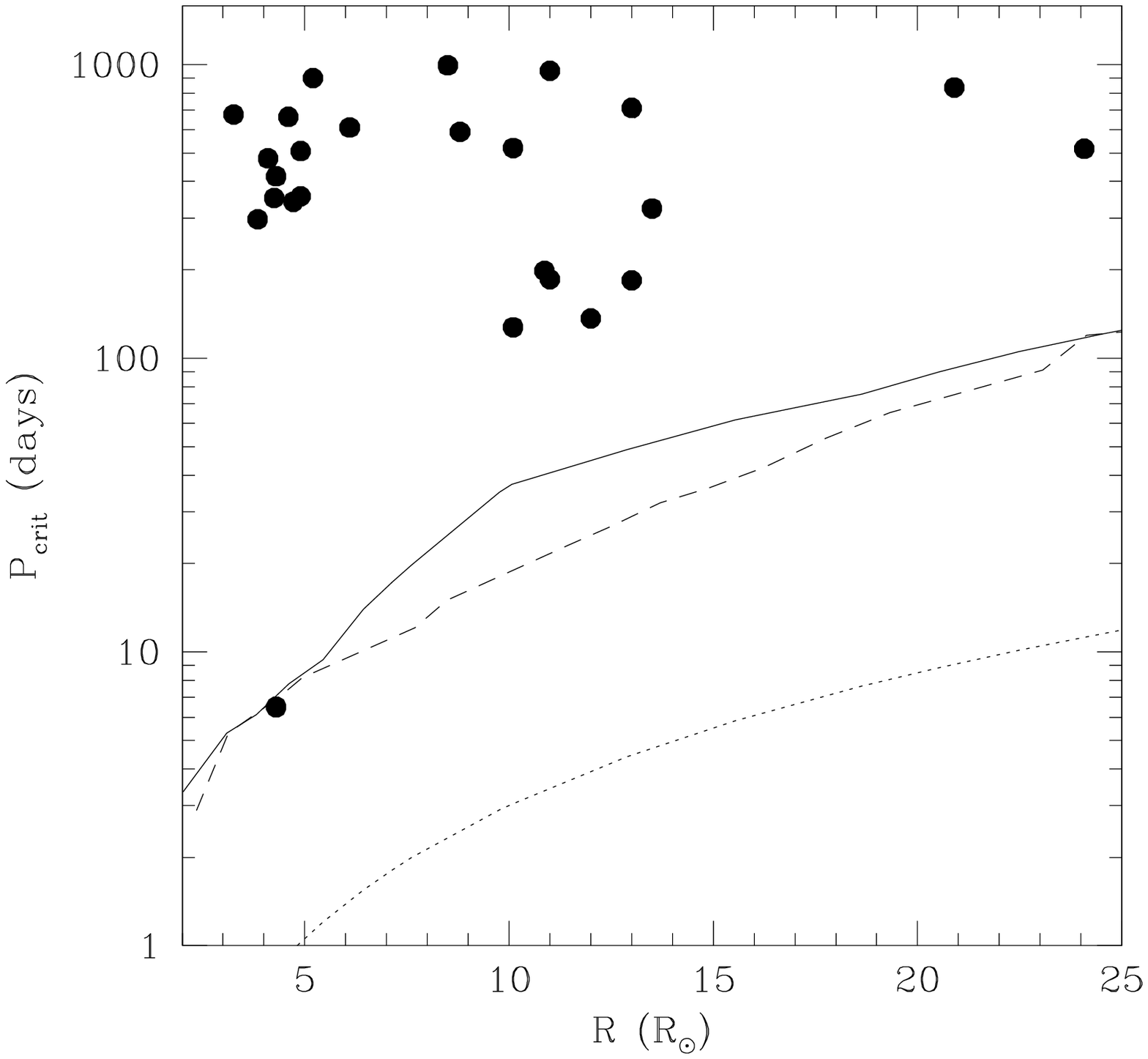]{The solid and dashed lines indicate the critical period for tidal
swallowing for $1.5 M_{\odot}$ and $2.5 M_{\odot}$ stars respectively, as a function of
the stellar radius. The dotted line indicates the Keplerian orbital period at the surface
of the evolving $1.5 M_{\odot}$ star. The filled circles are the observed orbital periods
for radial velocity planets around stars in the mass range 1.5--2.5 $M_{\odot}$.
\label{Swallow}}

\figcaption[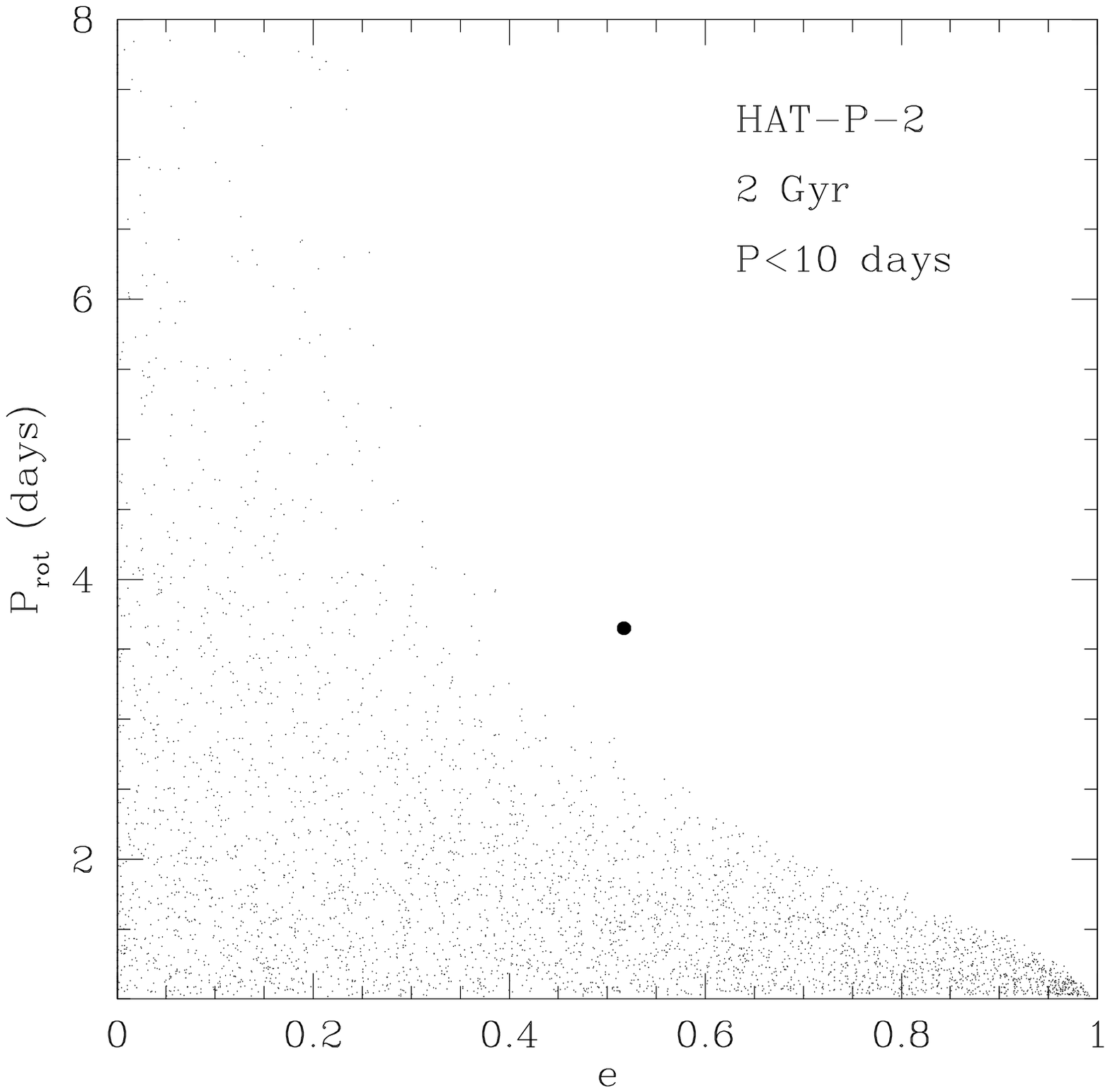]{The stellar rotation period versus planetary orbit eccentricity is shown for
the HAT-P-2 system parameters, after 2~Gyr of tidal evolution, and final orbital periods less
than 10~days. The solid point indicates the observed parameters, and the dots show the final state
for a variety of initial conditions, assuming the dissipation calibrations of \S~\ref{Synth}. The
fact that the simulations do not match the observed value indicates that rapid initial stellar spins
cannot stabilise the system in the face of strong stellar tidal dissipation.
\label{EOM}}

\figcaption[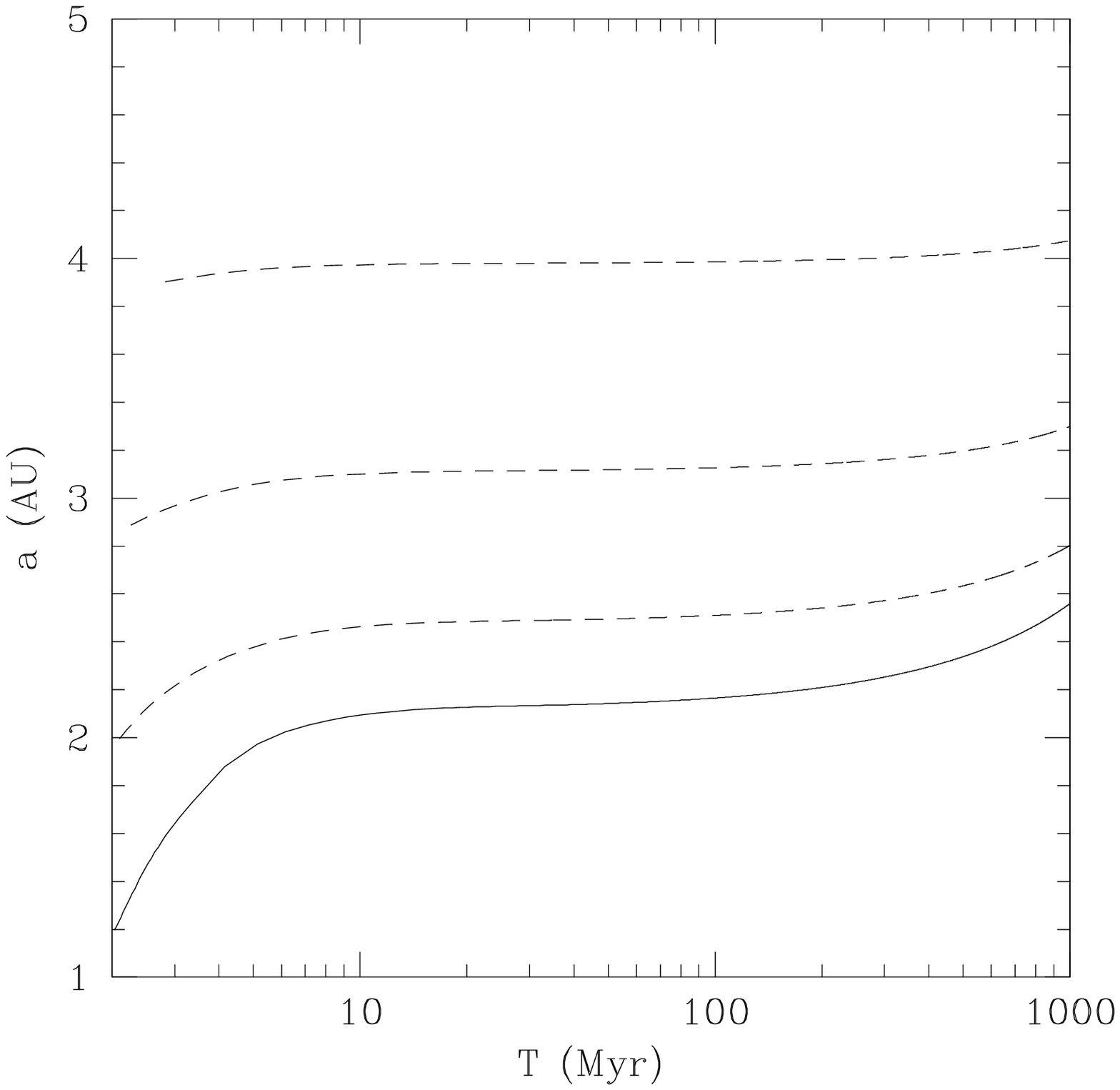]{The solid line indicates the expected orbital evolution of the planet WASP-33b,
driven by transfer of angular momentum from the spin of the host star to the planetary orbit. The
dashed lines are for other hypothetical systems in which WASP-33b began further out. We see that
the effects of tides decrease rapidly with distance. Note we have not included any effects on the
stellar rotation due to magnetic braking, mass loss or evolutionary changes in the star. These will
eventually become important and will likely change the evolutionary direction from outward to inward.
\label{WASP33}}

\figcaption[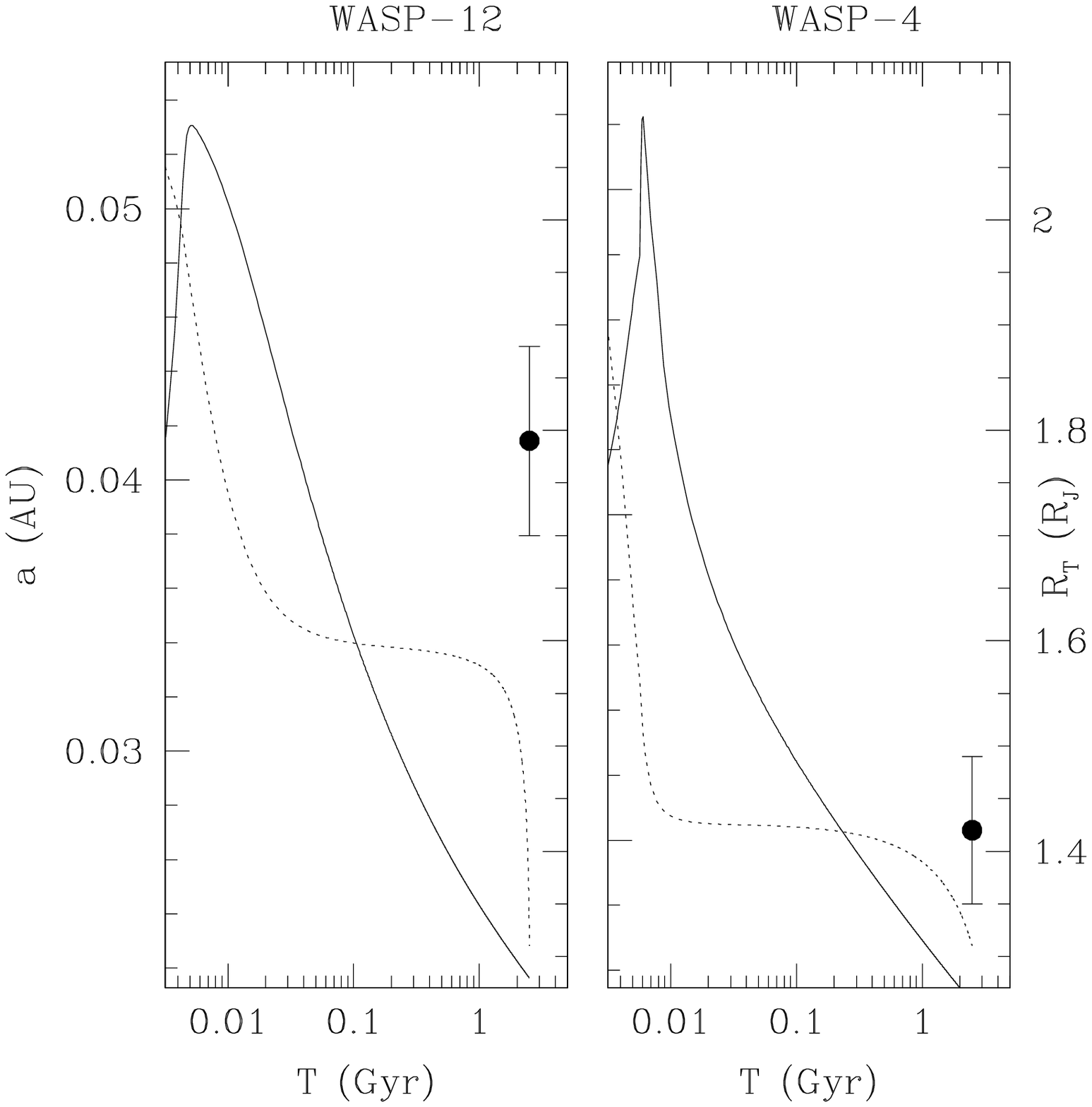]{ The left hand panel shows the evolution of the WASP-12b system and
the right hand panel shows the evolution of the WASP-4b system. In both cases the dotted curve
traces the semi-major axis evolution of the system, and corresponds to the left axis of the panel.
The solid line shows the evolution of the transit radius for the two systems, and corresponds
to the right axis of each panel. We see that planetary inflation does occur during tidal circulisation,
but that this occurs on timescales $\sim 10^7$ years, so that its influence is lost once the planet
has evolved for many planetary Kelvin-Helmholtz times, as is the case for these systems.
\label{twopanel}}

\figcaption[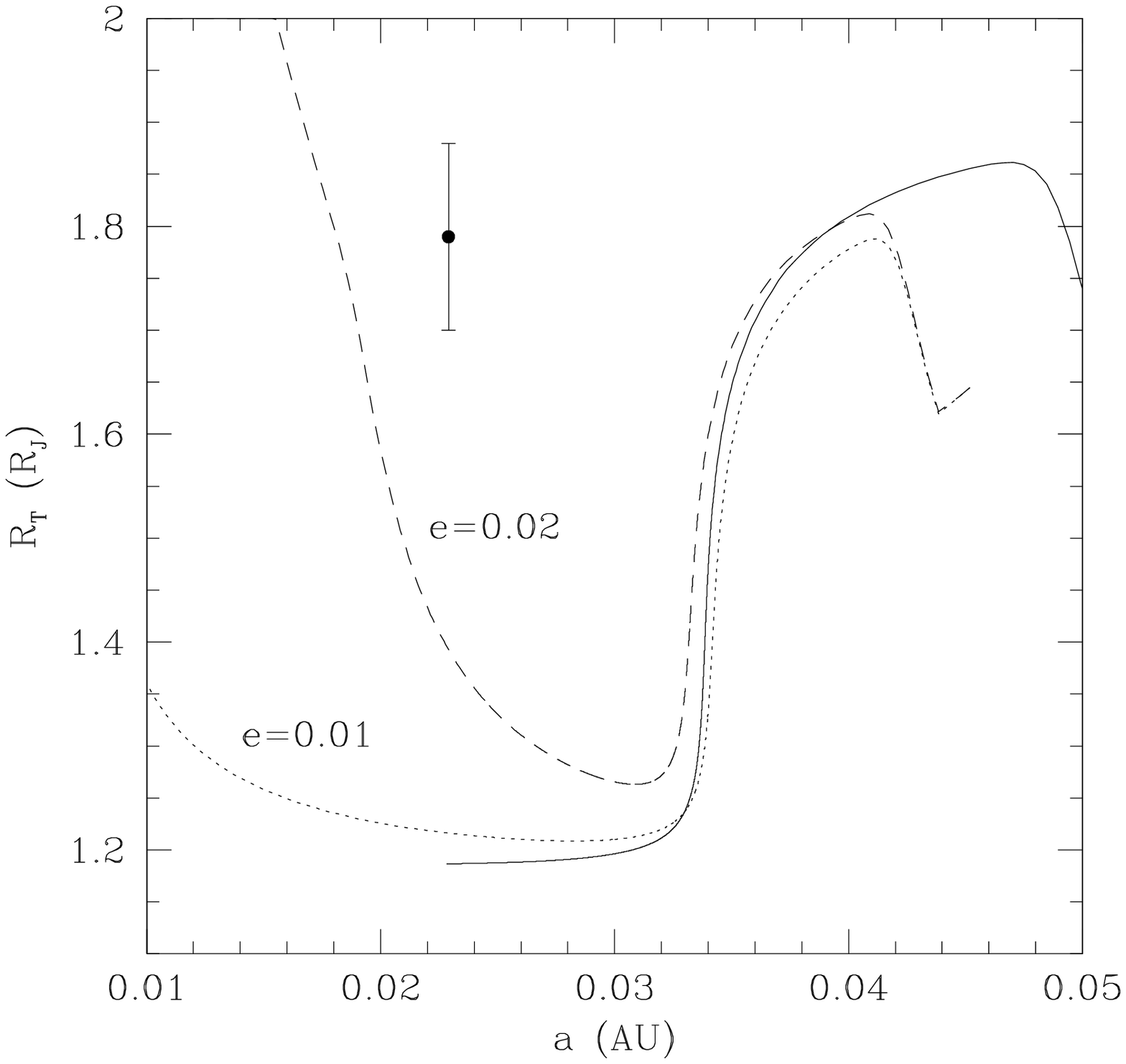]{ The solid curve shows the evolution of planetary transit radius
with semi-major axis during the inspiral of the WASP-12b system shown in the left
hand panel Figure~\ref{twopanel}. The dotted and dashed lines show the evolution of
the same system but where we have imposed an artificial lower limit to the eccentricity
of the orbit, as labelled. This results in an ongoing dissipation in the planet, which can
serve to inflate the planetary radius at late times.
\label{Floor}}

\figcaption[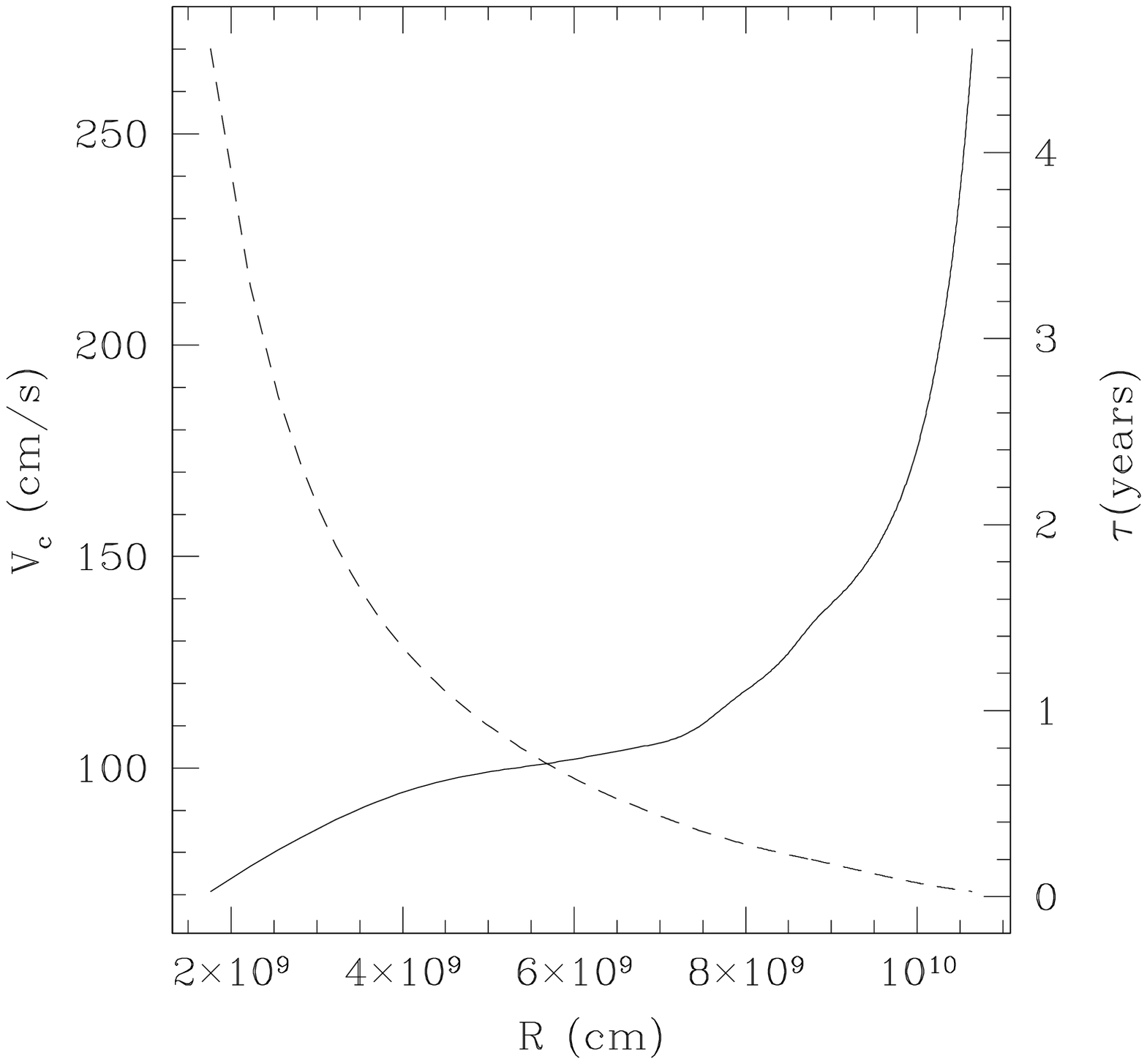]{ The solid curve is the profile of the convective velocity in the 
planet, given by the left hand axis. The dashed curve is the local eddy turnover time
$\tau = \ell_m/V_c$, where $\ell_m$ is taken to be a pressure scale height.
\label{vm}}

\figcaption[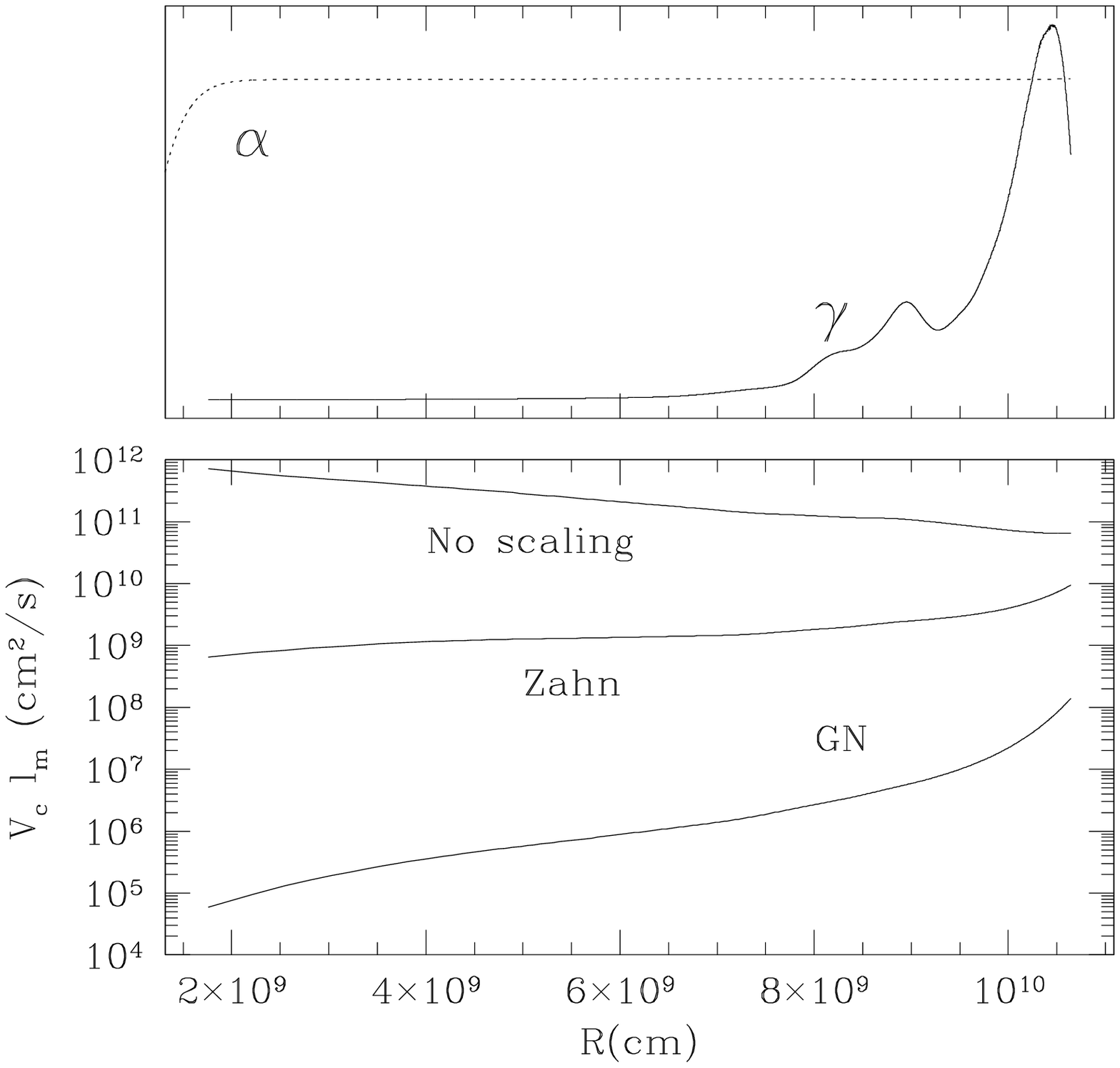]{The upper panel shows the radial profile of the function $\alpha$, which
represents the quadropolar distortion of the planet. We see this is pretty flat over most
of the volume. The function $\gamma$ is the weight function in EKH equation~(113), which 
describes how the tidal velocity field couples to the turbulent dissipation. We see this
favours the outer parts of the planet. In the lower panel, we show three functions representing
the local turbulent dissipation in the planet. The upper curve represents $V_c \ell_m$, while
the middle curve is $V_c \ell_m T/2 \tau$, where T(=3~days) is the tidal forcing period, and
$\tau$ is the local eddy turnover time. The lower curve is $V_c \ell_m (T/2 \pi \tau)^2$.
\label{wl}}

\figcaption[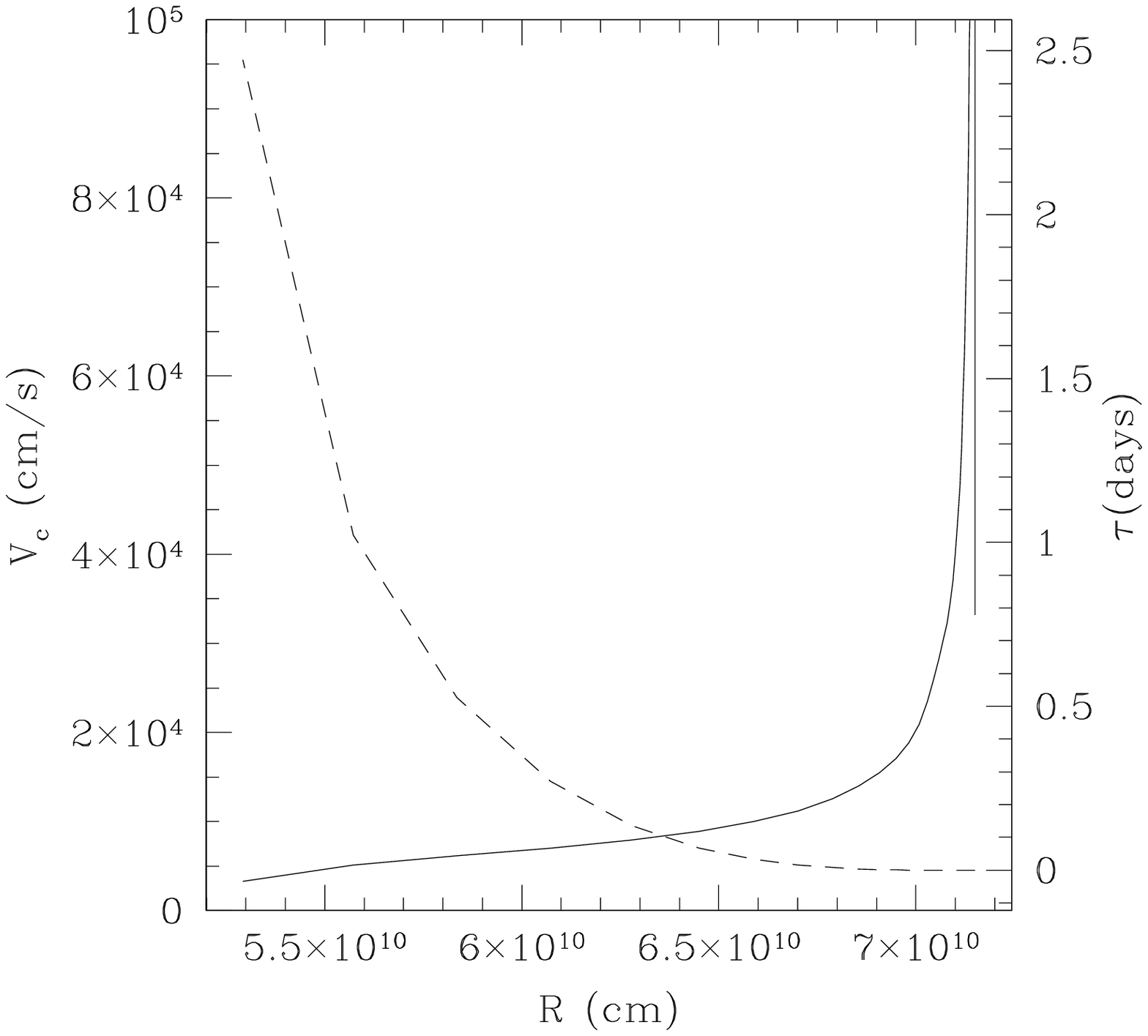]{The solid curve is the profile of the convective velocity in
the outer convection zone of a 5~Gyr old, $1 M_{\odot}$ stars. The dashed curve
indicates the eddy turnover time in this model, in units of days. Thus, close-in
planets with orbital periods of order days will match these timescales quite well.
\label{vms}}

\newpage

\appendix

\section{Internal Models and Turbulent Dissipation}
\label{Internal}

In order to relate our bulk dissipation to the underlying microphysics, we need to evaluate the various
equations for the quadropolar distortion given by EKH using a background from our planetary models. 
Figure~\ref{vm} shows the profile of convective velocity and large eddy turnover time $\tau$ for a 1~$M_J$ planet
at an age of $10^7$~years (we choose a young, hot model because this is the age where planets undergo the
bulk of their circularisation with our parameterisation). We see that $\tau$ has a value of order months over
the bulk of the planet, so that the coupling to the forcing period will not be particularly efficient.

Using the formalism from EKH, we can evaluate the profile of the quadropolar distortion $\alpha(r)$, and
the function $\gamma(r)$, which describes how the distortion couples to the microphysical dissipation. Figure~\ref{wl}
shows, in the top panel, the radial profile of these two functions. The distortion itself is relative flat throughout
most of the planet (whose density profile is approximately that of an n=1 polytrope), but the function $\gamma$ ascribes
a greater weight to the dissipation in the outer half of the planetary volume. In the lower panel of Figure~\ref{wl}
we show the radial profile of the turbulent viscosity $V_c \ell_m$ (where $\ell_m$ is the eddy scale, taken to be a
pressure scale height) in three cases. The upper curve is the full viscosity, with no corrections for inefficiency.
The two curves below that indicate the estimated local turbulent viscosity with linear and quadratic scalings for the
reduction in inefficiency, as suggested by Zahn or Goldreich \& Nicholson respectively. In these cases we have assumed
a forcing frequency of 3 days.

Following equation~(113) of EKH, one can integrate over the above profiles and infer a bulk viscosity. In the case of
a fully efficient coupling, we infer $\bar{\sigma}_p \sim 1.9 \times 10^{-4}$, which is $\sim 500$ times too strong. In
the case of a linear scaling, $\bar{\sigma}_p \sim 1.2 \times 10^{-5}$, which is still a factor of 30 too strong. For
a quadratic scaling, $\bar{\sigma}_p \sim 9.5 \times 10^{-8}$, which is only a factor of 3 weaker than our nominal
calibration. Thus, our inferred bulk dissipation is not very different from what one might expect on the basis of
the generally accepted picture of dissipation in turbulent convection.

In the case of the stellar dissipation, Figure~\ref{vms} shows the convective velocities and eddy turnover times
for the outer convection zone of a 5~Gyr, $1 M_{\odot}$ solar mass, taken from the EZ model (Paxton 2004). Repeating
the same exercise for the solar mass star yields $\bar{\sigma}_* \sim 7.8 \times 10^{-7}$, which is a factor $\sim 10$
larger than the inferred dissipation. In this case, the efficiency scalings do not make as large a difference,
as the turnover times are $\sim$ days, and so the reduction in efficiency only accounts for a factor of 2 change.
Thus, the inferred stellar value is also not too different from that expected on the basis of simple turbulent
dissipation, with some moderate inefficiency allowed.

\clearpage
\plotone{f1.ps}
\clearpage
\plotone{f2.ps}
\clearpage
\plotone{f3.ps}
\clearpage
\plotone{f4.ps}
\clearpage
\plotone{f5.ps}
\clearpage
\plotone{f6.ps}
\clearpage
\plotone{f7.ps}
\clearpage
\plotone{f8.ps}
\clearpage
\plotone{f9.ps}
\clearpage
\plotone{f10.ps}
\clearpage
\plotone{f11.ps}
\clearpage
\plotone{f12.ps}
\clearpage
\plotone{f13.ps}
\clearpage
\plotone{f14.ps}
\clearpage
\plotone{f15.ps}
\clearpage
\plotone{f16.ps}
\clearpage
\plotone{f17.ps}
\clearpage
\plotone{f18.ps}
\clearpage
\plotone{f19.ps}

\begin{references}
\reference{And} Anderson, D. R. et al., 2010, ApJ 709, 159
\reference{Bak} Bakos, G. et al., 2007, ApJ, 670, 826
\reference{BO} Barker, A. J. \& Ogilvie, G. I., 2010, arXiV:1001.4009
\reference{BHA} Barman, T. S., Hauschildt, P. H. \& Allard, F., 2005, ApJ, 632, 1132
\reference{BCAH} Baraffe, I., Chabrier, G., Allard, F. \& Hauschildt, P. H., 1998, A\&A, 337, 403
\reference{BCBAH} Baraffe, I., Chabrier, G., Barman, T., Allard, F. \& Hauschildt, P., 2003, A\&A, 407, 701
\reference{BBL} Batygin, K., Bodenheimer, P. \& Laughlin, G., 2009, ApJ, 704, L49
\reference{Bod} Bodenheimer, P. R., Lin, D. N. C., \& Mardling, R. A., 2001, ApJ, 548, 466
\reference{Bo10} Bowler, B. P., et al., 2010, ApJ, 709, 396
\reference{B00} Burrows, A., Guillot, T., Hubbard, W. B., Marley, M. S., Saumon,D., Lunine, J. I. \& Sudarsky, D., 2000, ApJ, 534, L97
\reference{BVM} Butler, R. P., Vogt, S., Marcy, G., Fischer, D., Henry, G. \& Apps, K., 2000, ApJ, 545, 504
\reference{Campo} Campo, C. et al., 2010, arXiV:1003.2763
\reference{CC} Collier Cameron, A., et al., 2010, arXiV:1004.4551
\reference{DaS} Da Silva, R. et al., 2005, A\&A, 446, 717
\reference{DD} Dobbs-Dixon, I., Lin, D. N. C. \& Mardling, R. A., 2004, ApJ, 610, 464
\reference{D10} Dunham, E., et al., 2010, ApJ, 713, L136
\reference{EKH} Eggleton, P., Kiseleva, L. \& Hut, P., 1998, ApJ, 499, 853
\reference{Padova} Girardi, L., Bresson, A., Bertelli, C. \& Chiosi, C., 2000, A\&AS, 141, 371
\reference{GN} Goldreich, P. \& Nicholson, P. D., 1977, Icarus, 30, 301
\reference{GS} Goldreich, P. \& Soter, S., 1966, Icarus, 5, 375
\reference{GD} Goodman, J. \& Dickson, E., 1998, ApJ, 507, 938
\reference{G09} Greenberg, R., 2009, ApJ, 698, L42
\reference{GBH} Guillot, T., Burrows, A., Hubbard, W. B., Lunine, J. I. \& Saumon, D., 1996, ApJ, 459, L35
\reference{HB} Hansen, B. \& Barman, T., 2007, ApJ, 671, 861
\reference{H09} Hebb, L. et al., 2009, ApJ, 693, 1920
\reference{H10} Hebb, L. et al., 2010, ApJ, 708, 224
\reference{Hell} Hellier, C., et al., 2009, Nature, 460, 1098
\reference{Husnoo} Husnoo, N., et al., 2010, arXiV:1004.1809
\reference{Hut} Hut, P, 1981, A\&A, 99, 126
\reference{IBS} Ibgui, L., Burrows, A. \& Spiegel, D. S., 2010, ApJ, 713, 751
\reference{J08} Jackson, B., Greenberg, R. \& Barnes, R., 2008a, ApJ, 678, 1396
\reference{J08b} Jackson, B., Greenberg, R. \& Barnes, R., 2008b, ApJ, 681, 1631
\reference{J09} Jackson, B., Barnes, R. \& Greenberg, R., 2009, ApJ, 698, 1357
\reference{JK} Johns-Krull, C. et al., 2008, ApJ, 677, 657
\reference{J06} Johnson, J., Marcy, G., Fischer, D., Henry, G., Wright, J., Isaacson, H. \& McCarthy, C., 2006, ApJ, 652, 1724
\reference{J07} Johnson, J. et al., 2007, ApJ, 665, 785
\reference{Josh} Joshi, Y. et al., 2009, MNRAS, 392, 1532
\reference{LAKH} Lainey, V., Arlot, J.-E., Karatekin, \"{O} \& van Hoolst, T., 2009, Nature, 459, 957
\reference{LCBL} Leconte, J., Chabrier, G., Baraffe, I. \& Levrard, H., 2010, arXiV:1004.0463
\reference{LM} Lopez-Morales, M., Coughlin, J. L., Sing, D. K., Burrows, A., Apai, D., Rogers, J. C. \& Spiegel, D. S., 2009, arXiV:0912.2359
\reference{LM2} Lovis, C. \& Mayor, M., 2007, A\&A, 472, 657
\reference{Marcy} Marcy, G. W., Butler, R. P., Williams, E., Bildsten, L., Graham, J. R., Ghez, A. M. \& Jernigan, J. G., 1997, ApJ, 481, 926
\reference{M07} Mardling, R., 2007, MNRAS, 382, 1768
\reference{MTR} Matsumura, S., Takeda, G. \& Rasio, F. A., 2008, ApJ, 686, L29
\reference{MM} Meibom, S. \& Mathieu, R. D., 2005, ApJ, 620, 970
\reference{MFJ} Miller, N., Fortney, J. \& Jackson, B., 2009, ApJ, 702, 1413
\reference{Nied} Niedzielski, A. et al., 2009, ApJ, 707, 768
\reference{Pax} Paxton, B., 2004, PASP, 116, 699
\reference{P99} Peale, S. J., 1999, ARA\&A, 37, 533
\reference{PL} Peale, S. J. \& Lee, M. H., 2002, Science, 298, 593
\reference{PBS} Penev, K., Barranco, J. \& Sasselov, D., 2009, ApJ, 705, 285
\reference{Pont} Pont, F., 2009, MNRAS, 396, 1789
\reference{RTLL} Rasio, F. A., Tout, C. A., Lubow, S. H. \& Livio, M., 1996, ApJ, 470, 1187
\reference{R09} Rauer, H. et al., 2009, A\&A, 506, 281
\reference{Sato} Sato, B., et al., 2008, PASJ, 60, 539
\reference{W08} Winn, J. et al., 2008, ApJ, 683, 1076
\reference{WS02} Witte, M. G. \& Savonije, G. J., 2002, A\&A, 386, 222
\reference{ZB} Zahn, J.-P. \& Bouchet, L., 1989, A\&A 223, 112
\reference{Z89} Zahn, J.-P, 1989, A\&A, 220, 112
\end{references}
\end{document}